\documentclass[longauth]{aa}
\usepackage{graphicx}
\usepackage{natbib}
\usepackage{scalerel}
\usepackage{comment}
\usepackage[table]{xcolor}
\usepackage{txfonts}
\usepackage[pdfencoding=auto,psdextra]{hyperref}
\hypersetup{
    colorlinks=true,
    linkcolor=blue,
    filecolor=magenta,      
    urlcolor=blue,
    citecolor=blue
}
\makeatletter
\renewcommand*\aa@pageof{, page \thepage{} of \pageref*{LastPage}}
\makeatother
\usepackage[utf8]{inputenc}
\usepackage[switch, modulo]{lineno}
\linenumbers
\usepackage{euclid}
\usepackage[normalem]{ulem}
\usepackage{tabularx}
\usepackage{rotating, booktabs}
\usepackage{inputenc}
\usepackage{geometry}
\usepackage{changepage}
\usepackage{color}
\usepackage{multirow}
\usepackage{lscape}            
\def\msun{\ensuremath{M_\odot}\xspace}
\def\Ha{{\ensuremath{\mathrm{H}\alpha}}\xspace}
\def\reff{\ensuremath{R_{\rm eff}}\xspace}

\begin{document} 

   \title{\Euclid: Early Release Observations $-$ The star-formation history of massive early-type galaxies in the Perseus cluster\thanks{This paper is published on behalf of the Euclid Consortium.}}

   \subtitle{}
											
\newcommand{\orcid}[1]{} 

\author{S.~Martocchia\orcid{0000-0001-7110-6775}\thanks{\email{silvia.martocchia@lam.fr}}\inst{\ref{aff1}}
\and A.~Boselli\orcid{0000-0002-9795-6433}\inst{\ref{aff1},\ref{aff2}}
\and J.-C.~Cuillandre\orcid{0000-0002-3263-8645}\inst{\ref{aff3}}
\and M.~Mondelin\orcid{0009-0004-5954-0930}\inst{\ref{aff3}}
\and M.~Bolzonella\orcid{0000-0003-3278-4607}\inst{\ref{aff4}}
\and C.~Tortora\orcid{0000-0001-7958-6531}\inst{\ref{aff5}}
\and M.~Fossati\orcid{0000-0002-9043-8764}\inst{\ref{aff6},\ref{aff7}}
\and C.~Maraston\orcid{0000-0001-7711-3677}\inst{\ref{aff8}}
\and P.~Amram\orcid{0000-0001-5657-4837}\inst{\ref{aff1}}
\and M.~Baes\orcid{0000-0002-3930-2757}\inst{\ref{aff9}}
\and S.~Boissier\orcid{0000-0002-9091-2366}\inst{\ref{aff1}}
\and M.~Boquien\orcid{0000-0003-0946-6176}\inst{\ref{aff10}}
\and H.~Bouy\orcid{0000-0002-7084-487X}\inst{\ref{aff11},\ref{aff12}}
\and F.~Durret\orcid{0000-0002-6991-4578}\inst{\ref{aff13}}
\and C.~M.~Gutierrez\orcid{0000-0001-7854-783X}\inst{\ref{aff14},\ref{aff15}}
\and M.~Kluge\orcid{0000-0002-9618-2552}\inst{\ref{aff16}}
\and Y.~Roehlly\orcid{0000-0001-8373-8702}\inst{\ref{aff1}}
\and T.~Saifollahi\orcid{0000-0002-9554-7660}\inst{\ref{aff17}}
\and M.~A.~Taylor\orcid{0000-0003-3009-4928}\inst{\ref{aff18}}
\and D.~Thomas\inst{\ref{aff8}}
\and T.~E.~Woods\orcid{0000-0003-1428-5775}\inst{\ref{aff19}}
\and G.~Zamorani\orcid{0000-0002-2318-301X}\inst{\ref{aff4}}
\and B.~Altieri\orcid{0000-0003-3936-0284}\inst{\ref{aff20}}
\and S.~Andreon\orcid{0000-0002-2041-8784}\inst{\ref{aff7}}
\and N.~Auricchio\orcid{0000-0003-4444-8651}\inst{\ref{aff4}}
\and C.~Baccigalupi\orcid{0000-0002-8211-1630}\inst{\ref{aff21},\ref{aff22},\ref{aff23},\ref{aff24}}
\and M.~Baldi\orcid{0000-0003-4145-1943}\inst{\ref{aff25},\ref{aff4},\ref{aff26}}
\and S.~Bardelli\orcid{0000-0002-8900-0298}\inst{\ref{aff4}}
\and P.~Battaglia\orcid{0000-0002-7337-5909}\inst{\ref{aff4}}
\and R.~Bender\orcid{0000-0001-7179-0626}\inst{\ref{aff16},\ref{aff27}}
\and A.~Biviano\orcid{0000-0002-0857-0732}\inst{\ref{aff22},\ref{aff21}}
\and E.~Branchini\orcid{0000-0002-0808-6908}\inst{\ref{aff28},\ref{aff29},\ref{aff7}}
\and M.~Brescia\orcid{0000-0001-9506-5680}\inst{\ref{aff30},\ref{aff5}}
\and S.~Camera\orcid{0000-0003-3399-3574}\inst{\ref{aff31},\ref{aff32},\ref{aff33}}
\and G.~Ca\~nas-Herrera\orcid{0000-0003-2796-2149}\inst{\ref{aff34},\ref{aff35}}
\and V.~Capobianco\orcid{0000-0002-3309-7692}\inst{\ref{aff33}}
\and C.~Carbone\orcid{0000-0003-0125-3563}\inst{\ref{aff36}}
\and J.~Carretero\orcid{0000-0002-3130-0204}\inst{\ref{aff37},\ref{aff38}}
\and M.~Castellano\orcid{0000-0001-9875-8263}\inst{\ref{aff39}}
\and G.~Castignani\orcid{0000-0001-6831-0687}\inst{\ref{aff4}}
\and S.~Cavuoti\orcid{0000-0002-3787-4196}\inst{\ref{aff5},\ref{aff40}}
\and K.~C.~Chambers\orcid{0000-0001-6965-7789}\inst{\ref{aff41}}
\and A.~Cimatti\inst{\ref{aff42}}
\and C.~Colodro-Conde\inst{\ref{aff14}}
\and G.~Congedo\orcid{0000-0003-2508-0046}\inst{\ref{aff34}}
\and C.~J.~Conselice\orcid{0000-0003-1949-7638}\inst{\ref{aff43}}
\and L.~Conversi\orcid{0000-0002-6710-8476}\inst{\ref{aff44},\ref{aff20}}
\and Y.~Copin\orcid{0000-0002-5317-7518}\inst{\ref{aff45}}
\and A.~Costille\inst{\ref{aff1}}
\and F.~Courbin\orcid{0000-0003-0758-6510}\inst{\ref{aff46},\ref{aff47},\ref{aff48}}
\and H.~M.~Courtois\orcid{0000-0003-0509-1776}\inst{\ref{aff49}}
\and H.~Degaudenzi\orcid{0000-0002-5887-6799}\inst{\ref{aff50}}
\and G.~De~Lucia\orcid{0000-0002-6220-9104}\inst{\ref{aff22}}
\and F.~Dubath\orcid{0000-0002-6533-2810}\inst{\ref{aff50}}
\and X.~Dupac\inst{\ref{aff20}}
\and S.~Escoffier\orcid{0000-0002-2847-7498}\inst{\ref{aff51}}
\and M.~Fabricius\orcid{0000-0002-7025-6058}\inst{\ref{aff16},\ref{aff27}}
\and M.~Farina\orcid{0000-0002-3089-7846}\inst{\ref{aff52}}
\and R.~Farinelli\inst{\ref{aff4}}
\and F.~Faustini\orcid{0000-0001-6274-5145}\inst{\ref{aff39},\ref{aff53}}
\and S.~Ferriol\inst{\ref{aff45}}
\and F.~Finelli\orcid{0000-0002-6694-3269}\inst{\ref{aff4},\ref{aff54}}
\and M.~Frailis\orcid{0000-0002-7400-2135}\inst{\ref{aff22}}
\and E.~Franceschi\orcid{0000-0002-0585-6591}\inst{\ref{aff4}}
\and P.~Franzetti\inst{\ref{aff36}}
\and M.~Fumana\orcid{0000-0001-6787-5950}\inst{\ref{aff36}}
\and S.~Galeotta\orcid{0000-0002-3748-5115}\inst{\ref{aff22}}
\and K.~George\orcid{0000-0002-1734-8455}\inst{\ref{aff55}}
\and B.~Gillis\orcid{0000-0002-4478-1270}\inst{\ref{aff34}}
\and C.~Giocoli\orcid{0000-0002-9590-7961}\inst{\ref{aff4},\ref{aff26}}
\and P.~G\'omez-Alvarez\orcid{0000-0002-8594-5358}\inst{\ref{aff56},\ref{aff20}}
\and J.~Gracia-Carpio\inst{\ref{aff16}}
\and A.~Grazian\orcid{0000-0002-5688-0663}\inst{\ref{aff57}}
\and F.~Grupp\inst{\ref{aff16},\ref{aff27}}
\and S.~V.~H.~Haugan\orcid{0000-0001-9648-7260}\inst{\ref{aff58}}
\and W.~Holmes\inst{\ref{aff59}}
\and I.~M.~Hook\orcid{0000-0002-2960-978X}\inst{\ref{aff60}}
\and F.~Hormuth\inst{\ref{aff61}}
\and A.~Hornstrup\orcid{0000-0002-3363-0936}\inst{\ref{aff62},\ref{aff63}}
\and K.~Jahnke\orcid{0000-0003-3804-2137}\inst{\ref{aff64}}
\and M.~Jhabvala\inst{\ref{aff65}}
\and B.~Joachimi\orcid{0000-0001-7494-1303}\inst{\ref{aff66}}
\and S.~Kermiche\orcid{0000-0002-0302-5735}\inst{\ref{aff51}}
\and A.~Kiessling\orcid{0000-0002-2590-1273}\inst{\ref{aff59}}
\and B.~Kubik\orcid{0009-0006-5823-4880}\inst{\ref{aff45}}
\and M.~K\"ummel\orcid{0000-0003-2791-2117}\inst{\ref{aff27}}
\and H.~Kurki-Suonio\orcid{0000-0002-4618-3063}\inst{\ref{aff67},\ref{aff68}}
\and A.~M.~C.~Le~Brun\orcid{0000-0002-0936-4594}\inst{\ref{aff69}}
\and D.~Le~Mignant\orcid{0000-0002-5339-5515}\inst{\ref{aff1}}
\and S.~Ligori\orcid{0000-0003-4172-4606}\inst{\ref{aff33}}
\and P.~B.~Lilje\orcid{0000-0003-4324-7794}\inst{\ref{aff58}}
\and V.~Lindholm\orcid{0000-0003-2317-5471}\inst{\ref{aff67},\ref{aff68}}
\and I.~Lloro\orcid{0000-0001-5966-1434}\inst{\ref{aff70}}
\and G.~Mainetti\orcid{0000-0003-2384-2377}\inst{\ref{aff71}}
\and D.~Maino\inst{\ref{aff72},\ref{aff36},\ref{aff73}}
\and O.~Mansutti\orcid{0000-0001-5758-4658}\inst{\ref{aff22}}
\and O.~Marggraf\orcid{0000-0001-7242-3852}\inst{\ref{aff74}}
\and M.~Martinelli\orcid{0000-0002-6943-7732}\inst{\ref{aff39},\ref{aff75}}
\and N.~Martinet\orcid{0000-0003-2786-7790}\inst{\ref{aff1}}
\and F.~Marulli\orcid{0000-0002-8850-0303}\inst{\ref{aff76},\ref{aff4},\ref{aff26}}
\and R.~J.~Massey\orcid{0000-0002-6085-3780}\inst{\ref{aff77}}
\and E.~Medinaceli\orcid{0000-0002-4040-7783}\inst{\ref{aff4}}
\and Y.~Mellier\inst{\ref{aff13},\ref{aff78}}
\and M.~Meneghetti\orcid{0000-0003-1225-7084}\inst{\ref{aff4},\ref{aff26}}
\and E.~Merlin\orcid{0000-0001-6870-8900}\inst{\ref{aff39}}
\and G.~Meylan\inst{\ref{aff79}}
\and A.~Mora\orcid{0000-0002-1922-8529}\inst{\ref{aff80}}
\and L.~Moscardini\orcid{0000-0002-3473-6716}\inst{\ref{aff76},\ref{aff4},\ref{aff26}}
\and C.~Neissner\orcid{0000-0001-8524-4968}\inst{\ref{aff81},\ref{aff38}}
\and S.-M.~Niemi\orcid{0009-0005-0247-0086}\inst{\ref{aff82}}
\and C.~Padilla\orcid{0000-0001-7951-0166}\inst{\ref{aff81}}
\and S.~Paltani\orcid{0000-0002-8108-9179}\inst{\ref{aff50}}
\and F.~Pasian\orcid{0000-0002-4869-3227}\inst{\ref{aff22}}
\and K.~Pedersen\inst{\ref{aff83}}
\and W.~J.~Percival\orcid{0000-0002-0644-5727}\inst{\ref{aff84},\ref{aff85},\ref{aff86}}
\and V.~Pettorino\orcid{0000-0002-4203-9320}\inst{\ref{aff82}}
\and S.~Pires\orcid{0000-0002-0249-2104}\inst{\ref{aff3}}
\and G.~Polenta\orcid{0000-0003-4067-9196}\inst{\ref{aff53}}
\and M.~Poncet\inst{\ref{aff87}}
\and L.~A.~Popa\inst{\ref{aff88}}
\and L.~Pozzetti\orcid{0000-0001-7085-0412}\inst{\ref{aff4}}
\and A.~Renzi\orcid{0000-0001-9856-1970}\inst{\ref{aff89},\ref{aff90}}
\and J.~Rhodes\orcid{0000-0002-4485-8549}\inst{\ref{aff59}}
\and G.~Riccio\inst{\ref{aff5}}
\and E.~Romelli\orcid{0000-0003-3069-9222}\inst{\ref{aff22}}
\and M.~Roncarelli\orcid{0000-0001-9587-7822}\inst{\ref{aff4}}
\and R.~Saglia\orcid{0000-0003-0378-7032}\inst{\ref{aff27},\ref{aff16}}
\and Z.~Sakr\orcid{0000-0002-4823-3757}\inst{\ref{aff91},\ref{aff92},\ref{aff93}}
\and A.~G.~S\'anchez\orcid{0000-0003-1198-831X}\inst{\ref{aff16}}
\and D.~Sapone\orcid{0000-0001-7089-4503}\inst{\ref{aff94}}
\and B.~Sartoris\orcid{0000-0003-1337-5269}\inst{\ref{aff27},\ref{aff22}}
\and P.~Schneider\orcid{0000-0001-8561-2679}\inst{\ref{aff74}}
\and A.~Secroun\orcid{0000-0003-0505-3710}\inst{\ref{aff51}}
\and G.~Seidel\orcid{0000-0003-2907-353X}\inst{\ref{aff64}}
\and S.~Serrano\orcid{0000-0002-0211-2861}\inst{\ref{aff95},\ref{aff96},\ref{aff97}}
\and E.~Sihvola\orcid{0000-0003-1804-7715}\inst{\ref{aff98}}
\and P.~Simon\inst{\ref{aff74}}
\and C.~Sirignano\orcid{0000-0002-0995-7146}\inst{\ref{aff89},\ref{aff90}}
\and G.~Sirri\orcid{0000-0003-2626-2853}\inst{\ref{aff26}}
\and J.~Steinwagner\orcid{0000-0001-7443-1047}\inst{\ref{aff16}}
\and P.~Tallada-Cresp\'{i}\orcid{0000-0002-1336-8328}\inst{\ref{aff37},\ref{aff38}}
\and A.~N.~Taylor\inst{\ref{aff34}}
\and I.~Tereno\orcid{0000-0002-4537-6218}\inst{\ref{aff99},\ref{aff100}}
\and N.~Tessore\orcid{0000-0002-9696-7931}\inst{\ref{aff101}}
\and S.~Toft\orcid{0000-0003-3631-7176}\inst{\ref{aff102},\ref{aff103}}
\and R.~Toledo-Moreo\orcid{0000-0002-2997-4859}\inst{\ref{aff104}}
\and F.~Torradeflot\orcid{0000-0003-1160-1517}\inst{\ref{aff38},\ref{aff37}}
\and I.~Tutusaus\orcid{0000-0002-3199-0399}\inst{\ref{aff97},\ref{aff95},\ref{aff92}}
\and L.~Valenziano\orcid{0000-0002-1170-0104}\inst{\ref{aff4},\ref{aff54}}
\and J.~Valiviita\orcid{0000-0001-6225-3693}\inst{\ref{aff67},\ref{aff68}}
\and T.~Vassallo\orcid{0000-0001-6512-6358}\inst{\ref{aff22}}
\and A.~Veropalumbo\orcid{0000-0003-2387-1194}\inst{\ref{aff7},\ref{aff29},\ref{aff28}}
\and Y.~Wang\orcid{0000-0002-4749-2984}\inst{\ref{aff105}}
\and J.~Weller\orcid{0000-0002-8282-2010}\inst{\ref{aff27},\ref{aff16}}
\and I.~A.~Zinchenko\orcid{0000-0002-2944-2449}\inst{\ref{aff106}}
\and E.~Zucca\orcid{0000-0002-5845-8132}\inst{\ref{aff4}}
\and J.~Garc\'ia-Bellido\orcid{0000-0002-9370-8360}\inst{\ref{aff107}}
\and J.~Mart\'{i}n-Fleitas\orcid{0000-0002-8594-569X}\inst{\ref{aff108}}
\and M.~Maturi\orcid{0000-0002-3517-2422}\inst{\ref{aff91},\ref{aff109}}
\and V.~Scottez\orcid{0009-0008-3864-940X}\inst{\ref{aff13},\ref{aff110}}}
										   
\institute{Aix-Marseille Universit\'e, CNRS, CNES, LAM, Marseille, France\label{aff1}
\and
INAF - Osservatorio Astronomico di Cagliari, Via della Scienza 5, 09047 Selargius (CA), Italy\label{aff2}
\and
Universit\'e Paris-Saclay, Universit\'e Paris Cit\'e, CEA, CNRS, AIM, 91191, Gif-sur-Yvette, France\label{aff3}
\and
INAF-Osservatorio di Astrofisica e Scienza dello Spazio di Bologna, Via Piero Gobetti 93/3, 40129 Bologna, Italy\label{aff4}
\and
INAF-Osservatorio Astronomico di Capodimonte, Via Moiariello 16, 80131 Napoli, Italy\label{aff5}
\and
Dipartimento di Fisica ``G. Occhialini", Universit\`a degli Studi di Milano Bicocca, Piazza della Scienza 3, 20126 Milano, Italy\label{aff6}
\and
INAF-Osservatorio Astronomico di Brera, Via Brera 28, 20122 Milano, Italy\label{aff7}
\and
Institute of Cosmology and Gravitation, University of Portsmouth, Portsmouth PO1 3FX, UK\label{aff8}
\and
Sterrenkundig Observatorium, Universiteit Gent, Krijgslaan 281 S9, 9000 Gent, Belgium\label{aff9}
\and
Universit\'e C\^{o}te d'Azur, Observatoire de la C\^{o}te d'Azur, CNRS, Laboratoire Lagrange, Bd de l'Observatoire, CS 34229, 06304 Nice cedex 4, France\label{aff10}
\and
Institut universitaire de France (IUF), 1 rue Descartes, 75231 PARIS CEDEX 05, France\label{aff11}
\and
Laboratoire d'Astrophysique de Bordeaux, CNRS and Universit\'e de Bordeaux, All\'ee Geoffroy St. Hilaire, 33165 Pessac, France\label{aff12}
\and
Institut d'Astrophysique de Paris, 98bis Boulevard Arago, 75014, Paris, France\label{aff13}
\and
Instituto de Astrof\'{\i}sica de Canarias, E-38205 La Laguna, Tenerife, Spain\label{aff14}
\and
Universidad de La Laguna, Dpto. Astrof\'\i sica, E-38206 La Laguna, Tenerife, Spain\label{aff15}
\and
Max Planck Institute for Extraterrestrial Physics, Giessenbachstr. 1, 85748 Garching, Germany\label{aff16}
\and
Universit\'e de Strasbourg, CNRS, Observatoire astronomique de Strasbourg, UMR 7550, 67000 Strasbourg, France\label{aff17}
\and
Department of Physics and Astronmy, University of Calgary, 2500 University Drive NW, Calgary, Alberta T2N 1N4, Canada\label{aff18}
\and
Department of Physics \& Astronomy, Allen Building, 30A Sifton Rd, University of Manitoba, Winnipeg MB R3T 2N2, Canada\label{aff19}
\and
ESAC/ESA, Camino Bajo del Castillo, s/n., Urb. Villafranca del Castillo, 28692 Villanueva de la Ca\~nada, Madrid, Spain\label{aff20}
\and
IFPU, Institute for Fundamental Physics of the Universe, via Beirut 2, 34151 Trieste, Italy\label{aff21}
\and
INAF-Osservatorio Astronomico di Trieste, Via G. B. Tiepolo 11, 34143 Trieste, Italy\label{aff22}
\and
INFN, Sezione di Trieste, Via Valerio 2, 34127 Trieste TS, Italy\label{aff23}
\and
SISSA, International School for Advanced Studies, Via Bonomea 265, 34136 Trieste TS, Italy\label{aff24}
\and
Dipartimento di Fisica e Astronomia, Universit\`a di Bologna, Via Gobetti 93/2, 40129 Bologna, Italy\label{aff25}
\and
INFN-Sezione di Bologna, Viale Berti Pichat 6/2, 40127 Bologna, Italy\label{aff26}
\and
Universit\"ats-Sternwarte M\"unchen, Fakult\"at f\"ur Physik, Ludwig-Maximilians-Universit\"at M\"unchen, Scheinerstr.~1, 81679 M\"unchen, Germany\label{aff27}
\and
Dipartimento di Fisica, Universit\`a di Genova, Via Dodecaneso 33, 16146, Genova, Italy\label{aff28}
\and
INFN-Sezione di Genova, Via Dodecaneso 33, 16146, Genova, Italy\label{aff29}
\and
Department of Physics "E. Pancini", University Federico II, Via Cinthia 6, 80126, Napoli, Italy\label{aff30}
\and
Dipartimento di Fisica, Universit\`a degli Studi di Torino, Via P. Giuria 1, 10125 Torino, Italy\label{aff31}
\and
INFN-Sezione di Torino, Via P. Giuria 1, 10125 Torino, Italy\label{aff32}
\and
INAF-Osservatorio Astrofisico di Torino, Via Osservatorio 20, 10025 Pino Torinese (TO), Italy\label{aff33}
\and
Institute for Astronomy, University of Edinburgh, Royal Observatory, Blackford Hill, Edinburgh EH9 3HJ, UK\label{aff34}
\and
Leiden Observatory, Leiden University, Einsteinweg 55, 2333 CC Leiden, The Netherlands\label{aff35}
\and
INAF-IASF Milano, Via Alfonso Corti 12, 20133 Milano, Italy\label{aff36}
\and
Centro de Investigaciones Energ\'eticas, Medioambientales y Tecnol\'ogicas (CIEMAT), Avenida Complutense 40, 28040 Madrid, Spain\label{aff37}
\and
Port d'Informaci\'{o} Cient\'{i}fica, Campus UAB, C. Albareda s/n, 08193 Bellaterra (Barcelona), Spain\label{aff38}
\and
INAF-Osservatorio Astronomico di Roma, Via Frascati 33, 00078 Monteporzio Catone, Italy\label{aff39}
\and
INFN section of Naples, Via Cinthia 6, 80126, Napoli, Italy\label{aff40}
\and
Institute for Astronomy, University of Hawaii, 2680 Woodlawn Drive, Honolulu, HI 96822, USA\label{aff41}
\and
Dipartimento di Fisica e Astronomia "Augusto Righi" - Alma Mater Studiorum Universit\`a di Bologna, Viale Berti Pichat 6/2, 40127 Bologna, Italy\label{aff42}
\and
Jodrell Bank Centre for Astrophysics, Department of Physics and Astronomy, University of Manchester, Oxford Road, Manchester M13 9PL, UK\label{aff43}
\and
European Space Agency/ESRIN, Largo Galileo Galilei 1, 00044 Frascati, Roma, Italy\label{aff44}
\and
Universit\'e Claude Bernard Lyon 1, CNRS/IN2P3, IP2I Lyon, UMR 5822, Villeurbanne, F-69100, France\label{aff45}
\and
Institut de Ci\`{e}ncies del Cosmos (ICCUB), Universitat de Barcelona (IEEC-UB), Mart\'{i} i Franqu\`{e}s 1, 08028 Barcelona, Spain\label{aff46}
\and
Instituci\'o Catalana de Recerca i Estudis Avan\c{c}ats (ICREA), Passeig de Llu\'{\i}s Companys 23, 08010 Barcelona, Spain\label{aff47}
\and
Institut de Ciencies de l'Espai (IEEC-CSIC), Campus UAB, Carrer de Can Magrans, s/n Cerdanyola del Vall\'es, 08193 Barcelona, Spain\label{aff48}
\and
UCB Lyon 1, CNRS/IN2P3, IUF, IP2I Lyon, 4 rue Enrico Fermi, 69622 Villeurbanne, France\label{aff49}
\and
Department of Astronomy, University of Geneva, ch. d'Ecogia 16, 1290 Versoix, Switzerland\label{aff50}
\and
Aix-Marseille Universit\'e, CNRS/IN2P3, CPPM, Marseille, France\label{aff51}
\and
INAF-Istituto di Astrofisica e Planetologia Spaziali, via del Fosso del Cavaliere, 100, 00100 Roma, Italy\label{aff52}
\and
Space Science Data Center, Italian Space Agency, via del Politecnico snc, 00133 Roma, Italy\label{aff53}
\and
INFN-Bologna, Via Irnerio 46, 40126 Bologna, Italy\label{aff54}
\and
University Observatory, LMU Faculty of Physics, Scheinerstr.~1, 81679 Munich, Germany\label{aff55}
\and
FRACTAL S.L.N.E., calle Tulip\'an 2, Portal 13 1A, 28231, Las Rozas de Madrid, Spain\label{aff56}
\and
INAF-Osservatorio Astronomico di Padova, Via dell'Osservatorio 5, 35122 Padova, Italy\label{aff57}
\and
Institute of Theoretical Astrophysics, University of Oslo, P.O. Box 1029 Blindern, 0315 Oslo, Norway\label{aff58}
\and
Jet Propulsion Laboratory, California Institute of Technology, 4800 Oak Grove Drive, Pasadena, CA, 91109, USA\label{aff59}
\and
Department of Physics, Lancaster University, Lancaster, LA1 4YB, UK\label{aff60}
\and
Felix Hormuth Engineering, Goethestr. 17, 69181 Leimen, Germany\label{aff61}
\and
Technical University of Denmark, Elektrovej 327, 2800 Kgs. Lyngby, Denmark\label{aff62}
\and
Cosmic Dawn Center (DAWN), Denmark\label{aff63}
\and
Max-Planck-Institut f\"ur Astronomie, K\"onigstuhl 17, 69117 Heidelberg, Germany\label{aff64}
\and
NASA Goddard Space Flight Center, Greenbelt, MD 20771, USA\label{aff65}
\and
Department of Physics and Astronomy, University College London, Gower Street, London WC1E 6BT, UK\label{aff66}
\and
Department of Physics, P.O. Box 64, University of Helsinki, 00014 Helsinki, Finland\label{aff67}
\and
Helsinki Institute of Physics, Gustaf H{\"a}llstr{\"o}min katu 2, University of Helsinki, 00014 Helsinki, Finland\label{aff68}
\and
Laboratoire d'etude de l'Univers et des phenomenes eXtremes, Observatoire de Paris, Universit\'e PSL, Sorbonne Universit\'e, CNRS, 92190 Meudon, France\label{aff69}
\and
SKAO, Jodrell Bank, Lower Withington, Macclesfield SK11 9FT, UK\label{aff70}
\and
Centre de Calcul de l'IN2P3/CNRS, 21 avenue Pierre de Coubertin 69627 Villeurbanne Cedex, France\label{aff71}
\and
Dipartimento di Fisica "Aldo Pontremoli", Universit\`a degli Studi di Milano, Via Celoria 16, 20133 Milano, Italy\label{aff72}
\and
INFN-Sezione di Milano, Via Celoria 16, 20133 Milano, Italy\label{aff73}
\and
Universit\"at Bonn, Argelander-Institut f\"ur Astronomie, Auf dem H\"ugel 71, 53121 Bonn, Germany\label{aff74}
\and
INFN-Sezione di Roma, Piazzale Aldo Moro, 2 - c/o Dipartimento di Fisica, Edificio G. Marconi, 00185 Roma, Italy\label{aff75}
\and
Dipartimento di Fisica e Astronomia "Augusto Righi" - Alma Mater Studiorum Universit\`a di Bologna, via Piero Gobetti 93/2, 40129 Bologna, Italy\label{aff76}
\and
Department of Physics, Institute for Computational Cosmology, Durham University, South Road, Durham, DH1 3LE, UK\label{aff77}
\and
Institut d'Astrophysique de Paris, UMR 7095, CNRS, and Sorbonne Universit\'e, 98 bis boulevard Arago, 75014 Paris, France\label{aff78}
\and
Institute of Physics, Laboratory of Astrophysics, Ecole Polytechnique F\'ed\'erale de Lausanne (EPFL), Observatoire de Sauverny, 1290 Versoix, Switzerland\label{aff79}
\and
Telespazio UK S.L. for European Space Agency (ESA), Camino bajo del Castillo, s/n, Urbanizacion Villafranca del Castillo, Villanueva de la Ca\~nada, 28692 Madrid, Spain\label{aff80}
\and
Institut de F\'{i}sica d'Altes Energies (IFAE), The Barcelona Institute of Science and Technology, Campus UAB, 08193 Bellaterra (Barcelona), Spain\label{aff81}
\and
European Space Agency/ESTEC, Keplerlaan 1, 2201 AZ Noordwijk, The Netherlands\label{aff82}
\and
DARK, Niels Bohr Institute, University of Copenhagen, Jagtvej 155, 2200 Copenhagen, Denmark\label{aff83}
\and
Waterloo Centre for Astrophysics, University of Waterloo, Waterloo, Ontario N2L 3G1, Canada\label{aff84}
\and
Department of Physics and Astronomy, University of Waterloo, Waterloo, Ontario N2L 3G1, Canada\label{aff85}
\and
Perimeter Institute for Theoretical Physics, Waterloo, Ontario N2L 2Y5, Canada\label{aff86}
\and
Centre National d'Etudes Spatiales -- Centre spatial de Toulouse, 18 avenue Edouard Belin, 31401 Toulouse Cedex 9, France\label{aff87}
\and
Institute of Space Science, Str. Atomistilor, nr. 409 M\u{a}gurele, Ilfov, 077125, Romania\label{aff88}
\and
Dipartimento di Fisica e Astronomia "G. Galilei", Universit\`a di Padova, Via Marzolo 8, 35131 Padova, Italy\label{aff89}
\and
INFN-Padova, Via Marzolo 8, 35131 Padova, Italy\label{aff90}
\and
Institut f\"ur Theoretische Physik, University of Heidelberg, Philosophenweg 16, 69120 Heidelberg, Germany\label{aff91}
\and
Institut de Recherche en Astrophysique et Plan\'etologie (IRAP), Universit\'e de Toulouse, CNRS, UPS, CNES, 14 Av. Edouard Belin, 31400 Toulouse, France\label{aff92}
\and
Universit\'e St Joseph; Faculty of Sciences, Beirut, Lebanon\label{aff93}
\and
Departamento de F\'isica, FCFM, Universidad de Chile, Blanco Encalada 2008, Santiago, Chile\label{aff94}
\and
Institut d'Estudis Espacials de Catalunya (IEEC),  Edifici RDIT, Campus UPC, 08860 Castelldefels, Barcelona, Spain\label{aff95}
\and
Satlantis, University Science Park, Sede Bld 48940, Leioa-Bilbao, Spain\label{aff96}
\and
Institute of Space Sciences (ICE, CSIC), Campus UAB, Carrer de Can Magrans, s/n, 08193 Barcelona, Spain\label{aff97}
\and
Department of Physics and Helsinki Institute of Physics, Gustaf H\"allstr\"omin katu 2, University of Helsinki, 00014 Helsinki, Finland\label{aff98}
\and
Departamento de F\'isica, Faculdade de Ci\^encias, Universidade de Lisboa, Edif\'icio C8, Campo Grande, PT1749-016 Lisboa, Portugal\label{aff99}
\and
Instituto de Astrof\'isica e Ci\^encias do Espa\c{c}o, Faculdade de Ci\^encias, Universidade de Lisboa, Tapada da Ajuda, 1349-018 Lisboa, Portugal\label{aff100}
\and
Mullard Space Science Laboratory, University College London, Holmbury St Mary, Dorking, Surrey RH5 6NT, UK\label{aff101}
\and
Cosmic Dawn Center (DAWN)\label{aff102}
\and
Niels Bohr Institute, University of Copenhagen, Jagtvej 128, 2200 Copenhagen, Denmark\label{aff103}
\and
Universidad Polit\'ecnica de Cartagena, Departamento de Electr\'onica y Tecnolog\'ia de Computadoras,  Plaza del Hospital 1, 30202 Cartagena, Spain\label{aff104}
\and
Caltech/IPAC, 1200 E. California Blvd., Pasadena, CA 91125, USA\label{aff105}
\and
Astronomisches Rechen-Institut, Zentrum f\"ur Astronomie der Universit\"at Heidelberg, M\"onchhofstr. 12-14, 69120 Heidelberg, Germany\label{aff106}
\and
Instituto de F\'isica Te\'orica UAM-CSIC, Campus de Cantoblanco, 28049 Madrid, Spain\label{aff107}
\and
Aurora Technology for European Space Agency (ESA), Camino bajo del Castillo, s/n, Urbanizacion Villafranca del Castillo, Villanueva de la Ca\~nada, 28692 Madrid, Spain\label{aff108}
\and
Zentrum f\"ur Astronomie, Universit\"at Heidelberg, Philosophenweg 12, 69120 Heidelberg, Germany\label{aff109}
\and
ICL, Junia, Universit\'e Catholique de Lille, LITL, 59000 Lille, France\label{aff110}}                  
   
   \date{Received XXX; accepted YYY}

  \abstract{The \Euclid\ Early Release Observations (ERO) programme targeted the Perseus galaxy cluster in its central region over 0.7\,deg$^2$. We combined the exceptional image quality and depth of the ERO-Perseus with FUV and NUV observations from GALEX and AstroSat/UVIT, as well as $ugriz\Ha$ data from MegaCam at the CFHT, to deliver FUV-to-NIR magnitudes of the 87 brightest galaxies within the Perseus cluster. We reconstructed the star-formation history (SFH) of 59 early-type galaxies (ETGs) within the sample, through the spectral energy distribution (SED) fitting code \texttt{CIGALE} and state-of-the-art stellar population (SP) models to reproduce the galactic UV emission from hot, old, low-mass stars (i.e. the UV upturn). In addition, for the six most massive ETGs in Perseus [stellar masses $\log_{10}(\Mstellar/\msun) \geq 10.3$], we analysed their spatially resolved SP and SFH through a radial SED fitting.
  In agreement with our previous work on Virgo ETGs, we found that (i) the majority of the analysed galaxies needs the presence of an UV upturn component to explain their FUV emission, with average temperatures $\langle T_{\rm UV}\rangle \simeq 33\,800$\,K; (ii) the ETGs of Perseus have grown their stellar masses quickly, with star-formation timescales $\tau\lesssim 1500$\,Myr. 
  We found that all ETGs in the sample have formed more than about 30\% of their stellar masses at $z\simeq5$, up to extreme fractions of $\simeq$100\%. At $z\simeq5$, the stellar masses of the most massive nearby ETGs, which have present-day stellar masses $\log_{10}(\Mstellar/\msun)\gtrsim 10.8$, are then found to be comparable to those of the red quiescent galaxies observed by JWST at similar redshifts ($z>4.6$).
  This study can be extended to ETGs in the 14\,000\,deg$^2$ extragalactic sky that will soon be observed by \Euclid, in combination with those from other major upcoming surveys (e.g. \textit{Rubin}/LSST), and UV observations, to ultimately assess whether the massive ETGs that we observe today represent the progeny of the massive high-$z$ JWST red quiescent galaxies.}

        
   \authorrunning{S. Martocchia et al.}
   \titlerunning{SFH of Perseus ETGs}
   \maketitle
\nolinenumbers


\section{Introduction}
\label{sec:intro}
The formation and evolution of red and quiescent massive galaxies at different cosmic epochs is still far from understood. At high redshifts, recent results from the {\it James Webb} Space Telescope (JWST) have shown that red galaxies are already very massive [stellar masses $\log_{10}(\Mstellar/\msun) > 10$] at $z>3$ (e.g. \citealt{baggen23, carnall23, carnall25, valentino23, labbe23, boyett24, glazebrook24, casey24, antwi-danso23, baker25b, baker25,  nanayakkara25, toni25}), and that they formed and quenched with extreme star-formation (SF) timescales (e.g. $\Delta t\in[100, 500]$\,Myr, representing here the duration of the SF burst for several parametrisations of the SFH, \citealt{degraaff25}). These observations are difficult to reconcile with predictions from hydrodynamical cosmological simulations (e.g. \citealt{lovell23, kimmig23}) and from galaxy formation and evolution scenarios (e.g. \citealt{oser10, delucia24, lagos25}), which rather predict a late formation of massive objects, posing challenges to the $\Lambda$-cold-dark-matter cosmological model (e.g. \citealt{menci22, boylan23}). 

To shed light on this issue, one approach is to reconstruct the star-formation history (SFH) of nearby, local massive early-type galaxies (ETGs), which we can observe in much more detail and with deeper observations. Indeed, such systems are red, old, and have little or no ongoing star-formation, resembling the high-$z$ galaxies observed by JWST. 
In addition, local massive ETGs are observed to be metal-rich and the most abundant in [$\alpha$/Fe] elements, which indicates that they must have formed rapidly (e.g. \citealt{worthey92, trager00a, thomas05, greene13}). As discussed in \cite{thomas05}, these results imply that quiescent massive galaxies with very high stellar masses should be present between redshifts 2 and 5, a prediction now confirmed by JWST. 
Nearby, local ETGs are indeed thought to form through a two-phase scenario \citep{oser10}. In the first phase, rapid in-situ star formation occurs, fueled by the infall of cold gas (at $z>2$), to explain the high metallicities and $\alpha$ abundances. To account for the difference in size among high and low redshift galaxies \citep{daddi05}, it is expected that a second phase is dominated by mass growth through dissipationless mergers (minor or major or both, occurring at later stages, $z<2$).
Connecting the populations of galaxies across different redshifts then provides fundamental insights into galaxy formation theories, as the properties of nearby ETGs are so similar to those of red and quiescent galaxies at high-$z$ that they might represent their descendants. 

A particular class of galaxies increasingly recognized as the direct descendants of massive red quiescent galaxies at high redshift are the so-called relic galaxies (e.g. \citealt{trujillo14, ferre-mateu17, yilidrim17, spiniello24, tortora25}). These systems are believed to have remained largely untouched by mergers over cosmic time, thus preserving the properties of their likely high-$z$ progenitors. They are characterized by very old stellar populations (SP, $\gtrsim 10$\,Gyr), compact sizes (\reff $\lesssim 2$\, kpc, \citealt{buitrago08}), and strong $\alpha$-enhancements ($[\alpha$/Fe]$\gtrsim 0.4$ dex; \citealt{trujillo14, ferre-mateu15, ferre-mateu17}). In some cases, they show only a single, red globular cluster (GC) population \citep{beasley18}, indicative of the absence of accretion from dwarf satellites.

The question we pose here is whether there exist ETGs in the local Universe (and if so, how many) whose SFH is consistent with that of the red quiescent galaxy population observed by JWST at high-$z$.
In this context, we started an observational campaign that aimed at reconstructing the SFH of massive ETGs in local galaxy clusters thanks to a multi-wavelength, radial analysis coupled with state-of-the-art SP models \citep{martocchia25}. Galaxy clusters provide ideal laboratories for the study of massive ETGs, as these galaxies are mainly found in such rich and dense environments (e.g. \citealt{dressler80}), suggesting that environment strongly shapes the evolution of galaxies (e.g.  \citealt{thomas10, labarbera12, boselli14, pasquali19}).
In \cite{martocchia25}, we reconstructed the SFH of seven massive ETGs within Virgo, our closest galaxy cluster, through a radial SED fitting analysis with the \texttt{CIGALE} code (Code Investigating GALaxy Evolution, e.g. \citealt{burgarella05, noll09, boquien19}). We recovered the properties of their spatially resolved SPs and found that these galaxies assembled their stellar mass extremely rapidly, consistently with what is found by JWST at high-$z$. 
In this study, we stressed the importance of a multi-wavelength approach in the reconstruction of the galaxies' SFHs. Originally thought to be simple, spheroidal, pressure-supported systems, most ETGs are now known to be fast rotators (FRs; e.g. \citealt{emsellem11}).They can also host gas in both molecular and ionised phases (e.g. \citealt{serra12, young14}), exhibit X-ray and radio emission from a central active galactic nucleus (AGN), and display strong UV emission from hot, old, low-mass stars in late evolutionary stages -- the so-called UV upturn phenomenon (e.g. \citealt{code79, oconnell99}).

Additionally, within ETGs, which usually host a limited amount of dust and gas, it is also imperative to have access to high angular resolution and deep data to detect small-scale and faint galactic components at different wavelengths. In these studies, \Euclid\ observations (\citealt{EuclidSkyOverview, EuclidSkyVIS, EuclidSkyNISP}) represent a major asset: the unprecedented data quality offered by \Euclid\ both in the visible and near-infrared (NIR) will, in the next few years, provide sharp and deep images of local massive ETGs in 14\,000\,deg$^2$ of the extragalactic sky. This will enable the characterisation of their morphology, the discovery of shells, tidal tails (indicative of merger events, e.g. \citealt{sola25}), as well as the presence of gas and dust structures, the study of the intracluster light \citep{EROPerseusICL}, and their star formation histories (\citealt{EP-Kovacic, EP-Abdurrouf, EP-Nersesian}). 
\Euclid\ observations, combined with surveys such as the Vera Rubin Observatory Legacy Survey of Space and Time (\textit{Rubin}/LSST) in the optical,  with the
Ultraviolet Near Infrared Optical Northern Survey (UNIONS, \citealt{ibata17}), with AstroSat in the far- and near- UV (and in the next decade, with the CASTOR and UVEX missions, \citealt{cote25} and \citealt{kulkarni21}), will ultimately be able to probe the properties of a statistically significant sample of nearby, massive ETGs in galaxy clusters of different masses and dynamical states. 

Here we use the \Euclid\ Early Release Observations (ERO, \citealt{EROcite}) of the Perseus cluster, in combination with other high spatial resolution multiwavelength observations of the same field, to reconstruct the SFH of the most massive ETGs in Perseus. The Perseus cluster is one of the most massive galaxy clusters in the local Universe, located at a distance of $\simeq 72$\,Mpc, dynamically older than Virgo. It has also been found to host the most promising local candidate of a relic galaxy, NGC\,1277 \citep{trujillo14}, together with many other compact ETGs. Despite being one of the most studied nearby clusters, the wide field of view (0.7\,deg$^2$, \citealt{EROPerseusOverview}), high spatial resolution, and low background of the \Euclid\ ERO provide the first high-resolution, full-cluster view of Perseus from the optical to the NIR.
We combine this unique dataset
with (i) images in the NUV and FUV from the UltraViolet Imaging Telescope (UVIT) on board the AstroSat satellite \citep{agrawal06}; (ii) NUV and FUV observations from GALEX; (iii) optical $ugriz\Ha$ images taken with MegaCam at the Canada-France-Hawaii-Telescope (CFHT). We provide integrated FUV-to-NIR magnitudes of the 87 brightest galaxies in Perseus. For 59 ETGs of the sample, we reconstruct their SFH from their integrated SEDs, while we report a radial SP analysis and SED fitting for six among the most massive ETGs in the sample.

This paper is structured as follows: Sect.~\ref{sec:obs} reports information on the data used, while Sect.~\ref{sec:analysis} describes the data analysis. 
In Sect.~\ref{sec:sed} we outline the SED fitting analysis and we report on the results in Sect.~\ref{sec:res}. Finally, we discuss in Sect.~\ref{sec:disc} and conclude in Sect.~\ref{sec:concl}.
Throughout the paper, we will use the AB magnitude reference system.

\section{Observations and data reduction}
\label{sec:obs}

\subsection{\Euclid\ ERO images}
\label{subsec:euclid}
The analysis presented in this paper is based on a set of multi-wavelength data of the Perseus cluster, based on the region imaged by the \Euclid\ ERO \citep{EROData, EROPerseusOverview}.
The Perseus cluster was imaged by \Euclid\ during the science verification phase in September 2023 \citep{EROData}, with a reference observing sequence (ROS, \citealt{Scaramella-EP1}). The ERO images consist of four ROSs, each one with four dithered exposures of 566\,s each in the $\IE$ filter (with effective wavelength $\lambda_{\rm eff} = 7171$\,\AA), and four dithered exposures of 87.2\,s each in the $\YE$, $\JE$, $\HE$ filters ($\lambda_{\rm eff} = 10\,809, 13\,673, 17\,714$\,\AA, \citealt{Schirmer-EP18, EuclidSkyVIS, EuclidSkyNISP}). For the specific passband curves, as well as for the data reduction procedures we refer to \cite{EROPerseusOverview}. The pixel sizes of the $\IE$ image is $\ang{;;0.1}$, while it is $\ang{;;0.3}$ for the NIR dataset ($\YE$, $\JE$, $\HE$); the full width at half maximum (FWHM) of the final $\IE$, $\YE$, $\JE$, $\HE$ science stacks measures $\ang{;;0.16}$, $\ang{;;0.48}$, $\ang{;;0.49}$, and $\ang{;;0.50}$, respectively. The dataset reaches a surface brightness limit of 30 (29.2) AB mag\,arcsec$^{-2}$ for the $\IE$ ($\YE$, $\JE$, $\HE$) band, excellent for the detection of low surface brightness (LSB) features such as shells and tidal tails formed during the gravitational interaction of galaxies with surrounding objects (e.g. \citealt{duc15, mancillas19, sola22}).

\subsection{UV imaging: GALEX and AstroSat/UVIT}
\label{subsec:uv}

In the UV regime, we downloaded already reduced images centred on NGC\,1275 from the GALEX\footnote{Through the Mikulski Archive for Space Telescopes (MAST) portal: \url{https://mast.stsci.edu/portal/Mashup/Clients/Mast/Portal.html}} and AstroSat/UVIT\footnote{\url{https://astrobrowse.issdc.gov.in/astro_archive/archive/Home.jsp}} archives. Table~\ref{tab:uv_tab} reports information on the collected UV datasets. We performed stacks of the long and short exposures of the GALEX images. The reached FWHM of the GALEX FUV and NUV images is $\simeq \ang{;;4.0}, \simeq \ang{;;5.3}$--$\ang{;;5.6}$, respectively \citep{morrissey07}\footnote{\url{https://asd.gsfc.nasa.gov/archive/galex/Documents/instrument_summary.html}}, while it is $\simeq \ang{;;1.3}$--$\ang{;;1.5}$, $\sim \ang{;;1.2}$--$\ang{;;1.4}$, for the AstroSat/UVIT images in the FUV and NUV bands, respectively \citep{tandon20}.
The reached surface brightness limits are 28.0 and 27.93 AB mag\,arcsec$^{-2}$ for the UVIT FUV and NUV dataset, while they are 26.4 and 26.1 AB mag\,arcsec$^{-2}$ for the GALEX FUV and NUV images, respectively. 

We used both GALEX and UVIT images to estimate the integrated fluxes and magnitudes of the Perseus galaxies in the FUV and NUV bands. We report the results for the GALEX bands in the main text of the paper (Sect.~\ref{subsec:integr}), as GALEX observations cover almost the entire \Euclid\ ERO footprint. In Appendix~\ref{app:comp_gal_uv}, we report on the comparison between the GALEX and UVIT magnitudes, see Sect.~\ref{subsec:integr} for more details. UVIT images, thanks to their unprecedented spatial resolution in the UV, were mainly used for the spatially resolved analysis of the SPs within the most massive ETGs of the Perseus cluster (see Sect.~\ref{subsec:rad_analys}).

\begin{table*}
\centering
\caption{Main information on the archival GALEX and AstroSat/UVIT images centred on the Perseus cluster used in this work.}
\begin{tabular}{ccccrc}
\hline\hline
\noalign{\smallskip}
Observatory & Filter & $\lambda_{\rm c}$ & Obs. ID & Exp. time & PI/Survey\\
            &        &  (\AA)        &         & (s)       &   \\
\hline
\noalign{\smallskip}
GALEX & FUV & 1524 & GI1\_098001\_A0426\_0002 & 14\,990 & R. O'Connell\\
GALEX & FUV & 1524 & NGA\_NGC1275 & 3345 & Nearby Galaxy Survey\\
GALEX & NUV & 2297 & GI1\_098001\_A0426\_0002 & 16\,249 & R. O'Connell\\
GALEX & NUV & 2297 & NGA\_NGC1275 & 3345 & Nearby Galaxy Survey\\
AstroSat & UVIT/BaF2 & 1541 & T01\_151T01\_9000000960 & 11\,099 & C. Varsha \\
AstroSat & UVIT/NUVB13 & 2447 & T01\_151T01\_9000000960 & 10\,966 & C. Varsha \\
\hline
\end{tabular}
\label{tab:uv_tab}
\end{table*}

\subsection{CFHT-MegaCam images}
\label{subsec:cfht}
We used optical images from the MegaCam at the CFHT \citep{EROPerseusOverview} which observed the Perseus cluster with a similar footprint of the \Euclid\ images in the $u$, $g$, $r$, $i$, $z$ third generation filters, at $\lambda_{\rm eff} \simeq 3682, 4784, 6397, 7695, 8987$\,\AA, respectively, as well as with an \Ha narrow-band filter (the `off' filter, CFHT ID 9604), centred on $\lambda_{\rm eff} = 6719$\,\AA, with a $\Delta\lambda=109$\,\AA. At the redshift of these galaxies, this filter includes the \Ha line together with the two \ion{N}{ii} lines at $\lambda=6548, 6583$\,\AA. Hereafter, we refer to the filter simply as \Ha, unless otherwise stated.
We refer to \cite{EROPerseusOverview} for details on the dataset information and data reduction process. The mean FWHM of the stacked $ugriz\Ha$ images is $\ang{;;1.46}$, $\ang{;;1.23}$, $\ang{;;0.79}$, $\ang{;;0.56}$, $\ang{;;0.60}$, $\ang{;;0.49}$, respectively. 

For this work, we generated stellar-continuum subtracted \Ha images, which are named ${\rm NET}\_ \Ha$ images hereafter. These images were used to check for contamination from young stars or recent SF in the Perseus galaxies. 
This step is particularly critical for ETGs where the emission of ionised gas is marginal compared to the emission of stars. As in \cite{boselli19,boselli22}, the stellar continuum was estimated by combining the $r$-band image with a $g-r$ colour map. Given the difference between the width and central transmission wavelength of the $r$ filter with respect to the narrow-band \Ha filter, the derivation of the stellar continuum depends on the spectral properties of the emitting source \citep{spector12}, thus on its colour. As in \cite{boselli18}, we used $\simeq 50\,000$ unsaturated stars with SDSS available spectroscopy to derive their magnitudes in the $g$, $r$, and \Ha filters and calibrate an empirical relation between the colour of the stars and the continuum. We obtained

\begin{equation}
    {\rm Cont\_}\Ha \, {\rm [mag]} =  r - 0.1123\,(g-r) - 0.0025.
\end{equation}

This normalisation was applied pixel by pixel, only where ${\rm S/N} > 1$, to avoid the introduction of any extra noise in the sky regions where no stellar continuum is present. We subtracted the normalised stellar continuum from the \Ha images pixel by pixel and we finally multiplied the continuum-subtracted image by the filter width ($\Delta\lambda=109$\,\AA), in order to obtain a flux from a flux density. We report the continuum-subtracted \Ha image of NGC\,1275 in the Appendix, Fig. \ref{fig:ha}.
The average unbinned sensitivity is $\Sigma(\Ha)\simeq 4 \times 10^{-17}$\,erg\,s$^{-1}$\,cm$^{-2}$\,arcsec$^{-2}$ at 1$\sigma$.
Since the contribution of the stellar continuum is derived using the $r$ and $g-r$ images, artefacts might be artificially created in those regions where the emission is strongly peaked and colour gradients are present, which typically happens in the core of bright ellipticals (e.g. \citealt{boselli19}, see also \citealt{martocchia25}). This might produce major effects in the resulting continuum-subtracted images. Any possible emission in these particular regions must be confirmed with spectroscopic data (see Sects.~\ref{subsec:suppl} and~\ref{subsec:gas} for more details on the analysis of the ionised gas emission).

\subsection{Supplementary data}
\label{subsec:suppl}
For the six targeted galaxies on which we will perform the radial SP analysis (see Sect.~\ref{subsec:rad_analys}), we checked their ionised nuclear gas emission. Thus, we downloaded SDSS optical spectra for NGC\,1270, NGC\,1281 and WISEAJ031922\_39+412545\_6; for NGC\,1277 and NGC\,1278, we downloaded optical spectra from the CfA survey \citep{falco99} to inspect for the presence of emission lines (\Ha, \hb, \ion{O}{iii}) in the nuclei, where the CFHT \Ha images might be contaminated by artefacts (see Sects.~\ref{subsec:cfht} and ~\ref{subsec:gas} for more details). For NGC\,1272, no optical spectrum is available in the literature, to the best of our knowledge.

We also downloaded images of the targeted galaxies in the infrared (IR) to estimate their dust masses (see Sect.~\ref{subsec:dust}). We downloaded {\it Spitzer}/IRAC level 2 images from the Spitzer Heritage Archive\footnote{\url{https://irsa.ipac.caltech.edu/applications/Spitzer/SHA/?__action=layout.showDropDown&}} (channels I1, I2, I3, and I4 where available, from $3.5$ to $8\,\mu$m) as well as WISE images from the IRSA/IPAC website in the bands W1, W2, W3, W4 (3.5, 4.6, 12, 22 $\mu$m)\footnote{\url{https://irsa.ipac.caltech.edu/applications/wise/?__action=layout.showDropDown&}}. We checked for far-IR (FIR) data of the targeted galaxies but only images from the Infrared Astronomical Satellite (IRAS) are available, with an angular resolution $> 100\arcsecond$ at $\lambda>60\,\mu$m, thus not usable for our purpose.

\begin{figure*}
\centering
\includegraphics[scale=0.45]{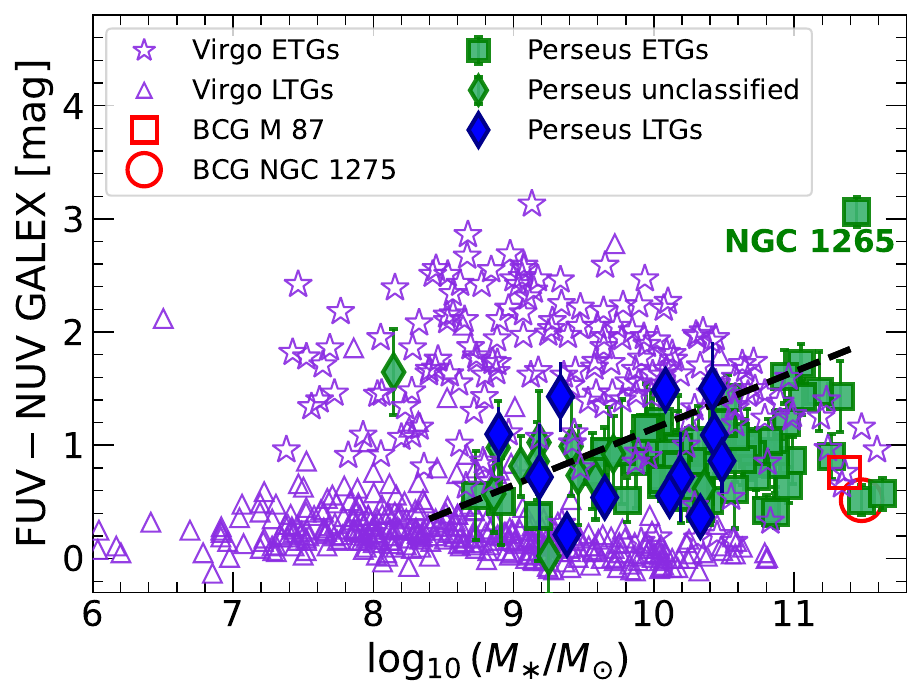}
\hspace{1cm}
\includegraphics[scale=0.45]{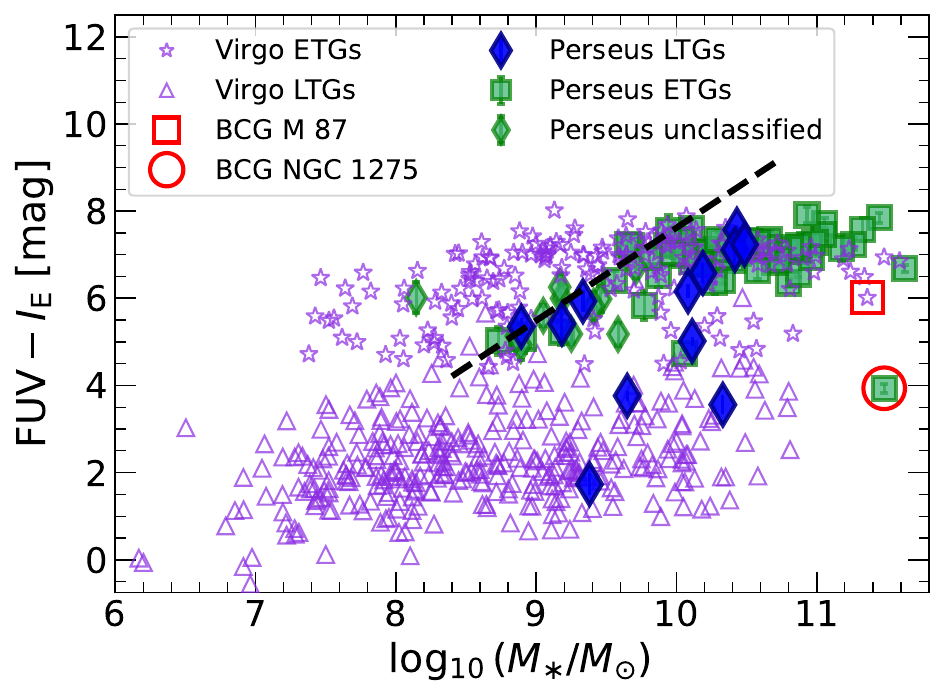}
\caption{${\rm FUV-NUV}$ (\emph{left}) and ${\rm FUV}-\IE$ (\emph{right}) colours as a function of stellar mass for the Perseus galaxies studied in this work and for the Virgo galaxies from the GUViCS data \citep{voyer14,boselli14}. Filled green squares (diamonds) indicate Perseus ETGs (unclassified) sources from \cite{meusinger20}. Filled blue diamonds are Perseus LTGs, while the open red circle represents the central galaxy of Perseus, namely NGC\,1275. Purple open triangles indicate Virgo LTGs, while Virgo ETGs are represented by purple open stars. M87, the central of Virgo, is also highlighted with an open red square. Black dashed lines are the magnitude limit curves for each corresponding colour for the Perseus observations. Circular apertures are defined in the FUV band starting with a radius equal to the \reff in \IE by \cite{EROPerseusOverview}. Apertures were adjusted when needed to fully encompass the FUV emission, and checked against the $\IE$ images to avoid contamination, as in \cite{voyer14}. We refer to Sect. \ref{subsec:integr} for more details.}
\label{fig:fuv_nuv_integr}
\end{figure*}

\section{Data analysis}
\label{sec:analysis}

\subsection{Integrated UV magnitudes of Perseus galaxies}
\label{subsec:integr}
We selected the 200 brightest galaxies in the $\IE$ magnitude from the catalogue of Perseus galaxy members by \cite{EROPerseusOverview}. We defined circle regions\footnote{We note that we used circle regions instead of ellipses, mainly, as the emission in the FUV and NUV bands of these galaxies has a circle-projected shape.} in the FUV band, by using a radius equal to the reported effective radius \reff in the $\IE$ band by \cite{EROPerseusOverview}. We note that this is a $\ensuremath{R_{\rm eff, maj}}$, thus it represents the semi-major axis of the isophote that includes 50\% of the flux and it is not circularised. We will refer to $\ensuremath{R_{\rm eff, maj}}$ as \reff hereafter. If the obtained circle region did not include all the FUV flux, the radius was increased or decreased manually, until the flux distribution of pixels homogeneously drops to zero. We also checked the aperture edges in the $\IE$ image, to make sure that the aperture was not extended too far and that it does not include a nearby source, as done in \cite{voyer14}. For the GALEX bands, we kept radii $\geq 0.5\,{\rm FWHM}$ ($>\ang{;;2.8}$). For two irregular galaxies, namely PGC012358 and UGC02665, we used an elliptical region. We then estimated the fluxes and magnitudes of such galaxies in each of the available bands (from FUV to $\HE$). Errors at 1$\sigma$ on fluxes and magnitudes are calculated as in \cite{martocchia25}, after masking all the images for foreground and background sources, and spikes produced by saturated stars. The background noise for each galaxy is estimated by averaging the flux of 1000 regions (with the same shape of each original region), each one generated with a random position in the ERO-Perseus field of view of each image.

We corrected the derived fluxes for the Milky Way extinction by assigning a colour excess $E(B-V)$ to each galaxy (see \citealt{EROPerseusOverview}) from the Planck 2013 dust opacity map \citep{planck_dust}, and by using the extinction curve by \cite{gordon23}, with $R_V=3.1$. For the extinction correction, we followed the prescriptions reported in \cite{EROPerseusOverview}, their Sect.~3.3 (see also \citealt{EROPerseusDGs}). We kept all sources with ${\rm S/N} \geq 3$ in both FUV and NUV bands. In Table~\ref{tab:fuv_nuv_integr}, we list the FUV and NUV magnitudes of all sources within the GALEX and UVIT field of view identified within the \Euclid\ footprint. In total, we identified 87 Perseus galaxies in the \Euclid\ footprint from the GALEX images (28 from UVIT) with a ${\rm S/N} \geq 3$ in all considered bands. Concerning the integrated study, we will refer to FUV and NUV magnitudes from GALEX, hereafter. 

\subsection{Colour comparison between Virgo and Perseus clusters}
\label{subsec:comp_virgo_pers}

We explore here UV and optical colours trends, as the ${\rm FUV}-\IE$ represents a pure comparison among hot stars, which are strong emitters in the FUV, and old cool stars, which are emitters in the optical, red bands. The ${\rm FUV-NUV}$ colour is instead an indication of the UV slope of the SED of galaxies. This colour has been largely considered in the literature as an empirical indicator of the UV upturn, because, when ${\rm FUV-NUV}<1$, the UV slope in the SED increases for decreasing wavelengths \citep{yi11}. However, this threshold depends on the assumed models and SFH, and it can occur that the UV emission from old stellar populations is also present when ${\rm FUV-NUV}>1$. Nevertheless, the  ${\rm FUV-NUV}$ colour is sensitive to different effects \citep{smithrussell12}, among which main sequence turn-off stars, due to strong metalline blanketing in the NUV \citep{donas07}. Here we make general considerations on the colour trends, and we refer to later Sections for the identification of the UV upturn in these galaxies in comparison with the models used in this work.
Figure~\ref{fig:fuv_nuv_integr} shows the ${\rm FUV-NUV}$ (left) and ${\rm FUV}-\IE$ (right) colours as a function of stellar mass for galaxies in the Perseus and Virgo clusters. 
Stellar masses of Perseus galaxies are from \cite{EROPerseusOverview}, which are calculated with a Chabrier IMF \citep{chabrier03} and the \cite{bc03} SP models. Data for the Virgo galaxies are from the GALEX Ultraviolet Virgo Cluster Survey (GUViCS, \citealt{voyer14} and \citealt{boselli14}).
To retrieve the $\IE$ magnitudes for the Virgo galaxies, we first converted the CFHT/MegaCam filters from the second (Virgo) to the third (Perseus) generation\footnote{We used the colour-magnitude transformations reported here: \url{https://www.cadc-ccda.hia-iha.nrc-cnrc.gc.ca/en/megapipe/docs/filt.html} together with the Virgo galaxy SDSS catalogue provided by \cite{cortese12}.}.
Next, the CFHT $i$ magnitudes were transformed to the $\IE$ band with a linear regression, as done in \cite{EROPerseusOverview}. 

Overall, in both panels, we observe that the Perseus ETGs and LTGs occupy a narrower part of the diagram compared to Virgo galaxies. We note that we are sampling the central regions of the Perseus cluster (one-fourth of the virial radius, \citealt{EROPerseusOverview}) while the Virgo GUViCS survey extends up to twice the virial radius, although it does not cover the entire extension of the cluster in the FUV. Therefore, we also examined Fig.~\ref{fig:fuv_nuv_integr} considering only the galaxies located within the inner one-fourth of the Virgo cluster’s virial radius. In this case, we note that the number of LTGs in Virgo is drastically reduced, as expected \citep{whitmore93}, showing that the bluest LTGs in Virgo are mainly further away from the cluster centre, where the density of the intracluster medium (ICM) is at its maximum and has likely transformed blue galaxies into red ones due to environmental processes \citep{boselli14}. 

Additionally, we observe that the Perseus and Virgo LTGs have quite different colours, with the Perseus LTGs being much redder in ${\rm FUV-NUV}$, comparable to those of ETGs in both clusters.\footnote{A similar behaviour is also observed when employing optical colours such as ${\rm g-i}$ colours.} Therefore, we visually inspected the Perseus LTGs in the \IE\ images. We found that only two of them are discs or spirals showing clumps and knots, indicative of the likely presence of gas and thus ongoing star formation.The remaining LTGs are either spirals or discs lacking a clear clumpy structure, and are likely gas-deficient systems (\citealt{mondelin25}, see their Fig. 3 for a snaphot of these galaxies).
To the best of our knowledge, no \ion{H}{i} data are available for these galaxies to perform a quantitative analysis of their gas deficiency (e.g. \citealt{boselli09}). The two gas-rich galaxies are UGC02665 and MCG+07-07-070, both of which are jellyfish galaxies known to be undergoing ram-pressure stripping (RPS) events \citep{roberts22, george25}. They also correspond to the LTGs in our sample with the bluest ${\rm FUV-NUV}$ colours, ${\rm FUV-NUV}<0.5$ in Fig. \ref{fig:fuv_nuv_integr}. It is therefore likely that the redder Perseus LTGs have been quenched due to RPS.
A comparison between RPS models and ${\rm FUV-NUV}$ colours in the Virgo cluster was presented by \citet[their Fig.~11]{boselli14}. The models predict a reddening of ${\rm FUV-NUV}$ depending on the lookback time, accompanied by quenching of star formation. The colour difference between Virgo and Perseus LTGs can then be explained by the higher efficiency of RPS in Perseus. The RPS efficiency depends on the ICM density and the velocity dispersion (${\rm RPS}=\rho_{{\rm ICM}}v^2$). Using velocity dispersions from \cite{boselligavazzi06} and \cite{kang24}, and ICM densities from \cite{urban11} and \cite{simionescu11} within radii $\in[0, 0.6]$ Mpc, we find that the RPS in Perseus is a factor of $\simeq$2 higher than that in Virgo.
\begin{table*}
\centering
\caption{Properties of the Perseus galaxies analysed radially. }
{\small
\begin{tabular}{lcrccc}
\hline\hline
ID & $\log_{10} (\Mstellar/\msun)^{(a)}$ & $\ensuremath{R_{\rm eff}}\xspace(\IE)$[kpc]$^{(a)}$ & Morph. Type$^{(b)}$ & AGN$^{(c)}$ & Fast/Slow Rot.$^{(d)}$\\
\hline
NGC\,1270 & 10.87 & 2.6 & E3 & Yes & FR\\
NGC\,1272 & 11.63 & 15.7 & S0 & Yes & SR\\
NGC\,1277 & 10.80 & 1.9 & S0 & Yes & FR\\
NGC\,1278 & 11.26 & 5.6 & E3 & Yes & SR\\
NGC\,1281 & 10.87 & 2.3 & S0 & Yes & FR\\
WISEAJ031922\_39+412545\_6 & 10.33 & 1.3 & S0 & No & ?\\
\hline
\label{tab:radial_data}
\end{tabular}
\tablebib{
\tablefoottext{a}{\citet{EROPerseusOverview}}
\tablefoottext{b}{\citet{meusinger20}}
\tablefoottext{c}{\citet{ferre-mateu15}, \citet{park17}, \citet{Saglia24}}
\tablefoottext{d}{\citet{raskutti14}, \citet{Saglia24}, \citet{emsellem13}, \citet{yildirim16}, \citet{comeron23}}
}
}
\end{table*}

\begin{figure*}
\centering
\includegraphics[scale=0.48]{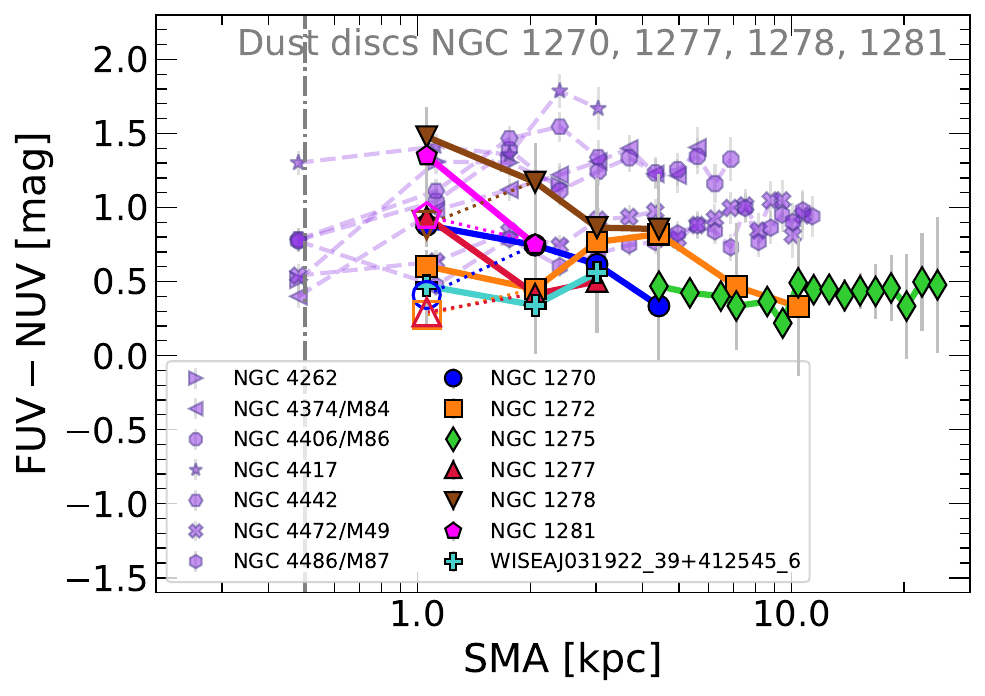}
\hspace{1cm}
\includegraphics[scale=0.48]{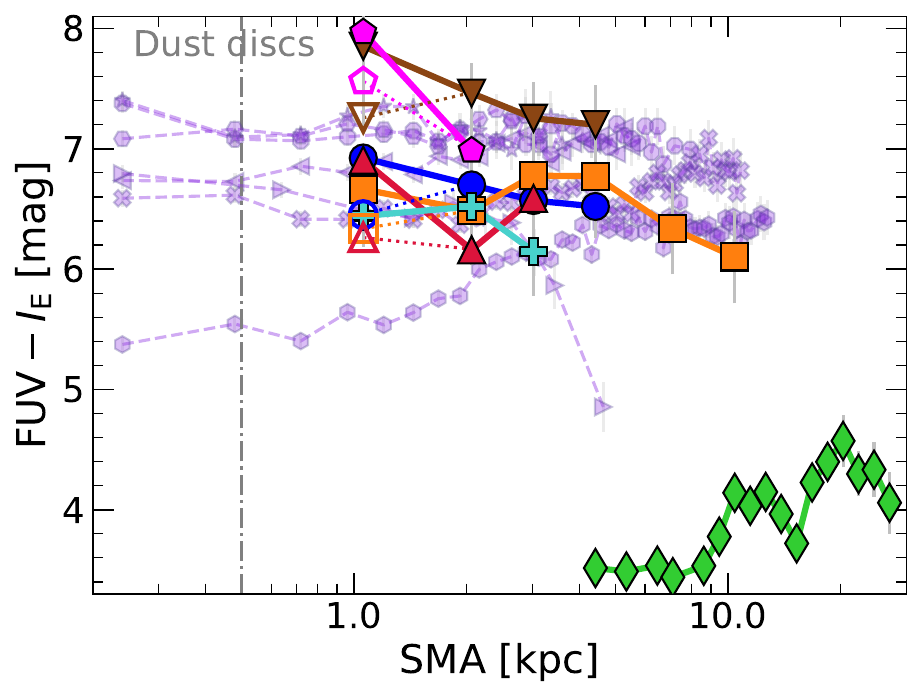}
\caption{${\rm FUV-NUV}$ (\emph{left}) and ${\rm FUV}-\IE$ (\emph{right}) colours as a function of the semimajor axis in kpc. Empty symbols indicate the values of the central colours calculated when a mask is not applied. The dotted lines connect the central unmasked regions to the successive regions. See Sect.~\ref{subsec:rad_analys} for more details. Perseus ETGs are indicated with solid lines, while Virgo ETGs from \cite{martocchia25} are represented by dashed lines and purple colours. The grey dash-dotted vertical line indicates the projected radius within which a disc of dust is observed in the $\IE$ images of NGC\,1270, NGC\,1277, NGC\,1278 and NGC\,1281, see Sect.~\ref{subsec:dust}.}
\label{fig:rad_UV}
\end{figure*}

Concerning the ETGs, we observe that at masses lower than $\log_{10}(\Mstellar/M_{\odot})\lesssim 10$, many Virgo ETGs are redder in ${\rm FUV-NUV}$ and in ${\rm FUV}-\IE$, while this population is apparently not observed in Perseus. This is most likely due to the fact that Perseus is further away than Virgo (72 Mpc vs. 16 Mpc), hence the redder, lower-mass (thus, lower luminosity) galaxies are not detected in the UV bands. This is shown by the magnitude limit curves plotted in Fig.~\ref{fig:fuv_nuv_integr} as black dashed lines for the Perseus observations\footnote{Calculated by interpolating the FUV, NUV and \IE versus stellar mass curves at $\log_{10}(\Mstellar/\msun)=8.7$ and $9.7$.}.
NGC\,1265 (labeled in Fig. \ref{fig:fuv_nuv_integr}), as well as the other ETG at $\log_{10}(\Mstellar/M_{\odot})\simeq 8.1$ are very red outliers. However, this is most likely due to the contamination of bright foreground stars that happens to lie at a projected distance of a few arcseconds from the centre of both galaxies. Both M\,87 and NGC\,1275 are quite blue, with UV slopes comparable to the bluest ETGs in both clusters, at similar stellar masses [$\log_{10}(\Mstellar/\msun) >11$]. This result is unsurprising, as both systems are the brightest cluster galaxies (BCGs) located at the centers of Virgo and Perseus, respectively, and therefore represent exceptional objects \citep{conselice01, boselli19}.

Also, Virgo and Perseus galaxies follow similar ${\rm FUV}-\IE$ colours, when taking into account the magnitude limits. 
Overall, ETGs are redder and more confined in ${\rm FUV}-\IE$ colours with respect to the LTGs, which span a wider range of ${\rm FUV}-\IE$ colours. This is true except for the two BCGs. NGC\,1275 is extremely blue with a ${\rm FUV}-\IE\simeq 4$, comparable to the LTGs in both clusters. M\,87 is also among the bluest of the ETGs sample. 
Again, we observe how LTGs in Virgo are bluer and much more numerous with respect to Perseus LTGs, at similar stellar masses, whose ${\rm FUV}-\IE$ colours are overall comparable to the ETGs. 
Among the LTGs of Perseus, three galaxies are very blue, with ${\rm FUV}-\IE<4$, namely UGC02665, WISEAJ032010\_12+412104\_2 and MCG+07-07-070, among which we find the two jellyfish galaxies mentioned above.
WISEAJ032010\_12+412104\_2 is also known as LEDA 12468, which has not been studied in detail so far. It appears as an elliptical galaxy, possibly with some faint discy gas emission, which needs to be confirmed with further analysis. 
Within this sample, we selected 59 ETGs in Perseus to perform a SED fitting analysis with \texttt{CIGALE} (Sect.~\ref{sec:sed}) by using the integrated fluxes reported in Tables~\ref{tab:fuv_nuv_integr} and~\ref{tab:optical_integr_galex}. We excluded NGC\,1275 for its cooling flow from this sample as well (see Fig. \ref{fig:ha}). We refer to this sample as the integrated sample, hereafter.

\subsection{Radial analysis}
\label{subsec:rad_analys}

For the radial analysis, we used the AstroSat/UVIT images in the FUV and NUV, given their higher spatial resolution (${\rm FWHM} \leq \ang{;;1.5}$) with respect to the GALEX images. We selected those ETGs (from Table~\ref{tab:fuv_nuv_integr}) that have a ${\rm S/N} \geq 7$ both in the UVIT FUV and NUV bands\footnote{We also tried to select galaxies with a ${\rm S/N} < 7$, however their radial ${\rm S/N}$ turned out to be too low ($< 2$) for a radial analysis.}. After this selection, we are left with 8 galaxies. 
Next, we followed the analysis described in \cite{martocchia25}, briefly summarised below; for further details, we refer the reader to that work.
Isophotal fitting was performed for each target galaxy on the $\IE$ images. The resulting isophotes were applied to all bands to derive fluxes in radial bins, using WCS coordinates to account for the differing image resolutions. Fits were carried out with the \texttt{ellipse} algorithm \citep{ellipse} implemented in \texttt{photutils} \citep{larry}. The position angle of successive ellipses was left free, as fixing it failed to reproduce galaxies with twisting isophotes. The isophote fit is then performed on consecutive ellipses of pre-defined semi-major axis (SMA), starting from a SMA $=\ang{;;3}$, and with a step of $\Delta$SMA $=\ang{;;3}$.

For the following radial analysis, we kept all galaxies that have at least three elliptical consecutive regions with ${\rm S/N} \geq3$ in the FUV band. The final list of the radially analysed ETGs is reported in Table~\ref{tab:radial_data}, along with their main properties. This consists of 7 galaxies, some of which are well-known and studied in the literature. Among these, we have 2 extended galaxies, NGC\,1272 and NGC\,1278, as well as 4 compact galaxies (\reff$\lesssim 3$\, kpc), including the well-studied relic galaxy NGC\,1277.
We refer to Appendix~\ref{app:ind_gal} for a more detailed description on each individual galaxy within the radial sample.

\begin{figure*}
\centering
\includegraphics[scale=0.47]{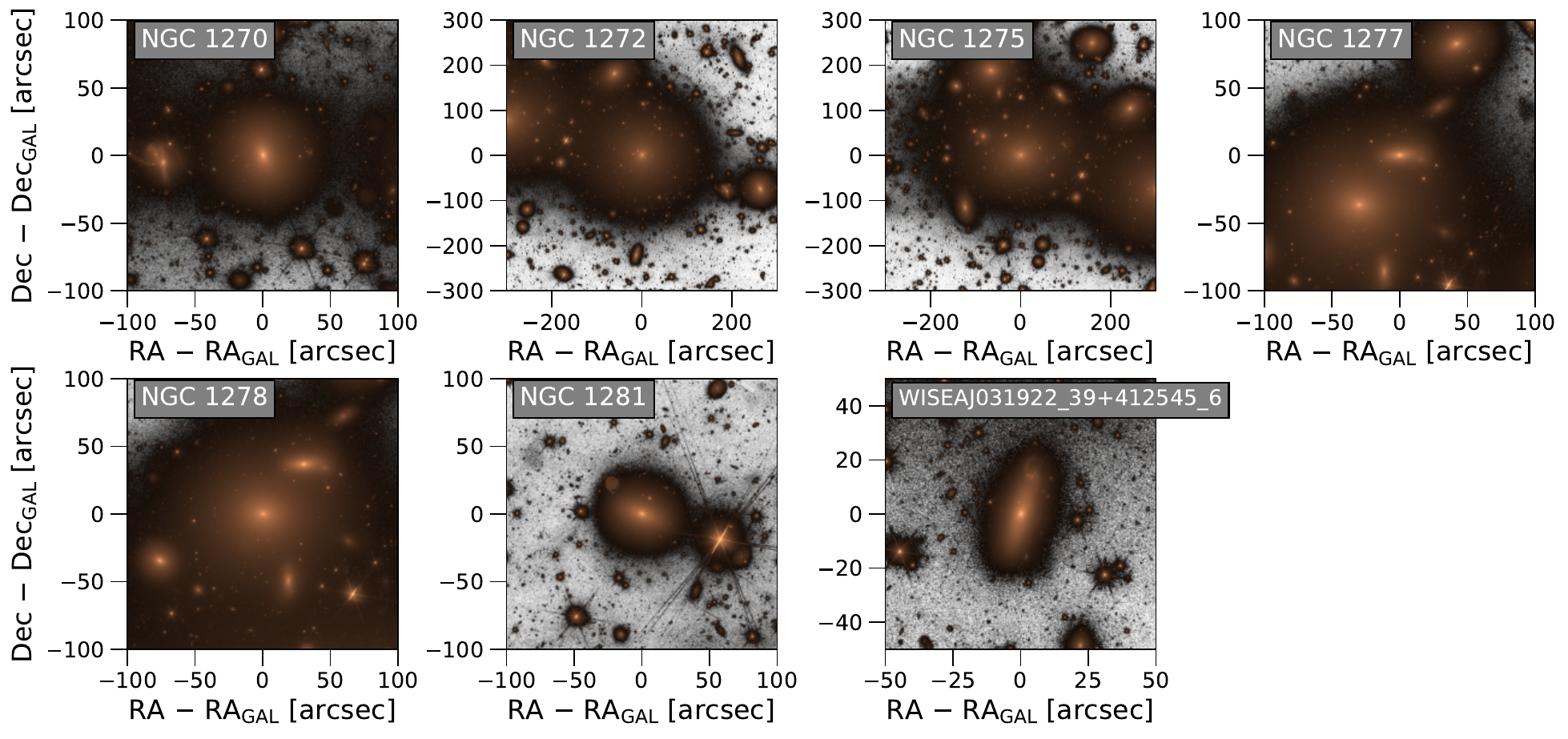}
\caption{Images of the ETGs in the radial sample in the $\IE$ band plotted with an asinh scale.}
\label{fig:sb_vis}
\end{figure*}

Based on the work by \cite{vigneron24}, we masked the inner regions of the BCG NGC\,1275 out to a radius of 2.6\,kpc from the centre of the galaxy, as the emission is dominated by the strong AGN and its nuclear activity. It is also well known that NGC\,1270, 1271, 1272, 1277, 1278, 1281 host an AGN in their centre, mostly in the form of peculiar very massive black holes (see Table \ref{tab:radial_data} and references therein). This was also recently confirmed by radio and X-ray observations \citep{santra07, park17, gendron-marsolais21, Saglia24}, while the AGN presence is not evident from the optical and IR parts of the nuclear spectra, where no significant emission lines are observed (e.g. \citealt{sakai12, park17}). These galaxies are indeed low-luminosity AGN, where their nuclear optical-IR spectra are dominated by the stellar activity. Hence, we do not expect a significant contribution from the AGN activity in the centre of these galaxies. However, as we are not able to disentangle the AGN UV emission from the one coming from the old stars, we masked the central region of these six galaxies (with $\ang{;;1}$ radius, corresponding to $\simeq$\,350 pc at the distance of Perseus) for both the integrated and radial analyses.

Figure~\ref{fig:rad_UV} shows the ${\rm FUV-NUV}$ (left) and ${\rm FUV}-\IE$ (right) colours as a function of the SMA of the considered ellipse region 
for both Perseus and Virgo galaxies (from \citealt{martocchia25})\footnote{For Virgo, we use that $\ang{;;1}$ corresponds to 80\,pc \citep{cantiello24}.}. 
Overall, the radial profiles are similar among the two clusters. As the right panel of Fig.~\ref{fig:rad_UV} shows, NGC\,1275 is very blue compared to the other ETGs, most likely due to the presence of young blue star clusters and young massive stars (\citealt{conselice01, canning14, tremblay15}).
For this reason, we discarded NGC\,1275 from the following SP analysis, as the presence of multi-phase gas and powerful AGN activity could hamper the interpretation of our results.  

\subsection{Galaxy morphology with \Euclid}
\label{subsec:morpho}

Thanks to the unprecedented depth and high angular resolution of the \Euclid\ observations, we were able to observe that four galaxies within the radial sample hosted a disc in their inner regions, at $0.1\lesssim$ SMA/$\reff \lesssim 0.2$, likely dusty discs or rings which might be either due to the presence of an AGN (e.g. \citealt{park17}) or similar to the nuclear discs observed in barred disc galaxies (e.g. \citealt{gadotti25}). The vertical lines in Fig.~\ref{fig:rad_UV} represent the projected distance from the centre within which the disc is observed. 
We will discuss the disc analysis in Sect.~\ref{subsec:dust}, in connection with the \texttt{CIGALE} SED fitting analysis.

In addition, we checked for LSB features in the surroundings of the galaxies in the radial sample, which might represent an indication of recent major or minor mergers, as well as gravitational interactions that galaxies underwent (e.g. \citealt{mancillas19, sola22, sola25}). 
With a surface brightness limit of 30\,mag\,arcsec$^{-2}$ in the $\IE$ band, the \Euclid\ ERO dataset represents a significant advance over previous wide-field surveys. For comparison, SDSS typically reaches $\simeq$26\,mag\,arcsec$^{-2}$, DES achieves $\simeq$27\,mag\,arcsec$^{-2}$, and the Subaru Hyper Suprime-Cam survey extends to 28\,mag\,arcsec$^{-2}$. The unprecedented depth of \Euclid\ thus opens new avenues for the study of faint, extended LSB structures in galaxies, stellar halos, and the diffuse intergalactic medium.
Figure~\ref{fig:sb_vis} shows the images of the ETGs in the radial sample in the $\IE$ band, plotted as in \cite{sola22} to enhance the contrast. Our inspection reveals the lack of LSB structures. However, in the particular case of the ERO Perseus, the LSB detection might be hampered by two main issues: (i) the first is the observed crowding of elliptical galaxies towards the centre of the cluster, (ii) the second is the presence of the Galactic cirri emission, with a median surface brightness of 27.5\,mag\,arcsec$^{-2}$ \citep{EROPerseusOverview}, due to the low Galactic latitude of the Perseus cluster.

\begin{figure*}
\centering
\includegraphics[scale=0.3]{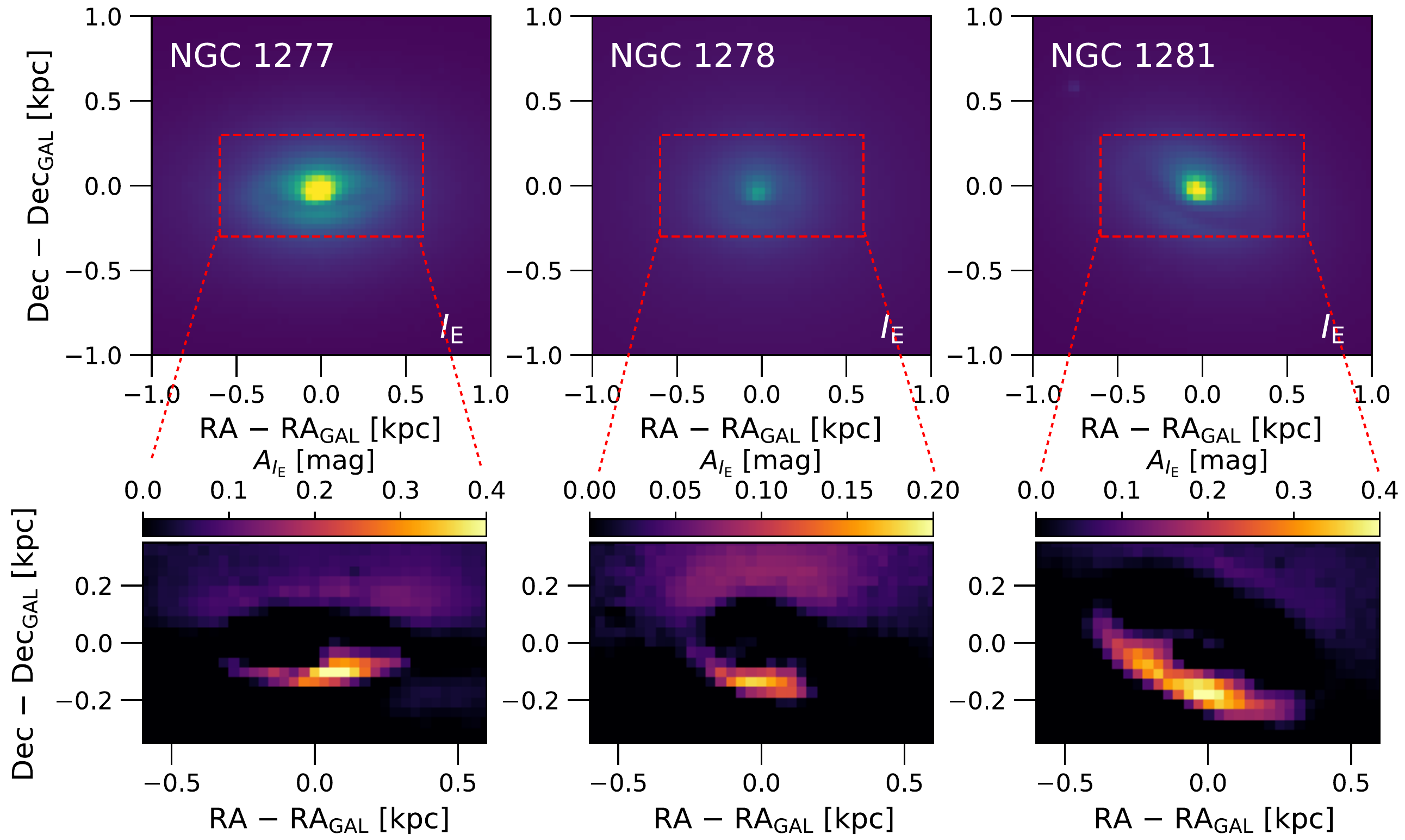}
\caption{\emph{Top panels}: \IE\ images zoomed in the inner regions of NGC\,1277, NGC\,1278, and NGC\,1281, from left to right. \emph{Bottom panels}: extinction maps in the $\IE$ filter, $A_{\IE}$, constructed from the top panels. North is up and east is to the left.}
\label{fig:discs}
\end{figure*}

\section{Spectral energy distribution fitting}
\label{sec:sed}

We constructed FUV-to-NIR SEDs, with the FUV, NUV, $ugriz$, \IE, \YE, \JE, \HE magnitudes. For the integrated sample of galaxies we used the FUV and NUV magnitudes from GALEX, while, for the radial sample, we constructed the SEDs as a function of the SMAs for each galaxy in Table~\ref{tab:radial_data}, by using the FUV and NUV magnitudes from AstroSat/UVIT. To reconstruct the SFH of the target galaxies, we then fit the SEDs with the code \texttt{CIGALE} \citep{boquien19}, similarly to what was done in our previous work \citep{martocchia25}. More details on the adopted modules and fitting procedures are reported in Sect.~\ref{subsec:cigale}. For the redshift, we used the average $z=0.0167$ of the cluster, from \cite{EROPerseusOverview}. For the six galaxies within the radial sample, the best-fit SEDs are shown in Fig. \ref{fig:res_sed_radial}.

\subsection{Adopted CIGALE modules}
\label{subsec:cigale}
Table~\ref{tab:cigale_p} reports on the adopted \texttt{CIGALE} modules and parameters used to fit the galaxies' SEDs. 
We model the SFH with the \texttt{sfhdelayed} module, namely in the form SFR$(t)\propto t/\tau^{2} \exp(-t/\tau)$, where $t$ is the SFH range and $\tau$ represents the e-folding timescale of SF. 

Concerning the stellar models, we used the \texttt{uvupturn} module \citep{martocchia25} to take into account the UV emission of evolved stars (the phenomenon of the UV upturn), based on the models by \cite{maraston05}. The models include a UV upturn in the form of an old, hot stellar component at certain defined temperatures, as in \cite{mt00} and \cite{lecras16}, which simulates the evolution of post-main sequence (PMS) stars. This is not modelled using stellar tracks, thus the UV emission might come from hot horizontal branch stars as well as from post asymptotic giant branch, AGB-manqu\'e, and other PMS evolutionary phases.  We can then explore the main stellar parameters responsible for UV upturn stars regardless of their evolutionary path. We adopt a \cite{salpeter55} IMF, the only IMF currently available for the upturn models. We note that the Salpeter IMF is bottom-heavier than Kroupa/Chabrier IMFs and is commonly associated with more massive ETGs \citep{labarbera13, tortora13, spiniello14}\footnote{We note that evidence for IMF variations has been found in massive elliptical galaxies, with IMFs bottom-heavier in their centre with respect to their periphery \citep{martin-navarro15, vandokkum17}. In a follow-up study, we will constrain the radial variation of the IMF in these galaxies and to extend the models to multiple IMF prescriptions, allowing for radially varying IMFs where supported by the data.}.
However, in the figures below and throughout the paper, we will report stellar masses calculated with a Chabrier IMF, in line with the results by \cite{EROPerseusOverview}. We note indeed that there is only a difference in stellar mass by a factor of $\simeq$0.25 dex downward with respect to a Salpeter IMF \citep{cimatti08, tortora09, tortora10}.
We refer to \cite{martocchia25} for more detail about the models and the \texttt{CIGALE} parameters. 
The UV emission in the used stellar models is degenerate with the emission produced by recent episodes of SF and young stars. To break this degeneracy, we used the CFHT \Ha images in order to constrain the contribution of the young stellar population component. We measured the \Ha emission of each galaxy and thus their star-formation rate (SFR), and we used this as a constraint in the SED fitting. See Sect.~\ref{subsec:gas} for more details. In Appendix~\ref{subsec:noup_sedfit}, we report on additional tests to constrain the possible contribution of a young stellar component to explain the UV emission of the target galaxies.

Regarding dust attenuation, we used the \texttt{dustatt\_modified\_starburst} module, based on the \cite{calzetti00} law. Additionally, we used the \texttt{themis} module to model dust emission, although our dataset only spans a wavelength range in the IR up to $\simeq 1.7\,\mu$m, where dust emission is expected to be weak. We based our parameter space choice on the work by \cite{nersesian19} which investigated the properties of dust in more than 800 nearby galaxies. We report additional considerations and tests on the dust modelling in Sect.~\ref{subsec:dust}.

\begin{figure*}
\centering
\includegraphics[scale=0.6]{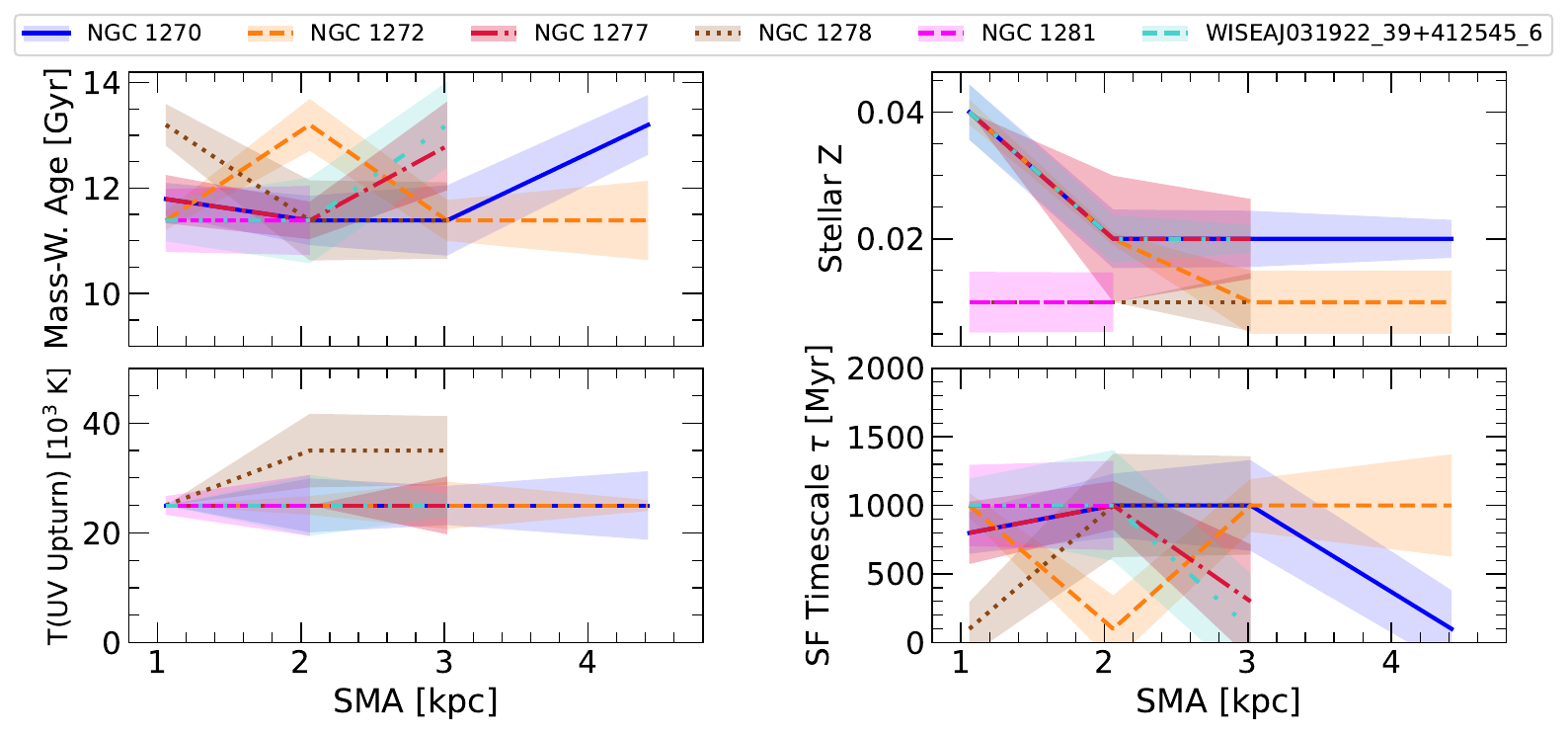}
\caption{SED fitting results of galaxy properties for the radial sample of Perseus ETGs. From the upper left to the lower right: mass-weighted age, stellar metallicity, temperature of the UV upturn, and SF timescale as a function of the SMA/\reff. Shaded areas represent 1$\sigma$ errors.}
\label{fig:res_sed}
\end{figure*}

\subsection{Ionised gas emission}
\label{subsec:gas}
The continuum-subtracted CFHT-\Ha images of the target galaxies, as well as the FUV and NUV images, show a lack of structured regions such as clumps or knots, which demonstrates the absence of star forming complexes in both the integrated and radial samples. 
Additionally, for the radial sample, we checked the available optical spectra (see Sect.~\ref{subsec:suppl}) to identify the presence of emission lines within the galaxies' nuclei. We did not find any emission line in any selected galaxy. 
Hence, as done in \cite{martocchia25}, we used the continuum-subtracted CFHT-\Ha images to measure an upper limit on the SFR in each galaxy from their \Ha emission. We estimated the surface brightness limit in \Ha by taking the standard deviation of the flux of 1000 regions (with the same shape as the original region), each one generated with a random position in the field of view. We then converted the \Ha flux at 5$\sigma$ into SFRs (\citealt{kennicutt98}, Eq. 2) and we included this rest-frame property in the SED fitting procedure through \texttt{CIGALE} as upper limit.

\subsection{Dust modelling and AGN contamination}
\label{subsec:dust}

In order to understand how to treat the dust absorption and emission in the fit, we inspected the $\IE$ images and looked for filamentary structures in absorption, for galaxies in the radial and integrated sample. None is observed. The only visible dust structure is found in a disc-shape in the inner 0.5\,kpc$^{2}$ close to the centre of NGC\,1270, 1277, 1278, and 1281. To get an estimation of the stellar extinction, we constructed a 2D model of the stellar distribution for each galaxy in the sample, on the $\IE$ images through the same isophote analysis presented in Sect.~\ref{subsec:rad_analys}. We then subtracted the stellar 2D model from the original image and we estimated the extinction coefficient in the $\IE$ band, $A_{\IE}$ (as in \citealt{kulkarni14,boselli22, martocchia25}). The subtraction of the 2D model does not find any significant disc for NGC\,1270, for which the dust disc is very weak and also barely visible in the \IE\ image. The visible dust discs and extinction maps are reported in Fig.~\ref{fig:discs} for NGC\,1277 (left), NGC\,1278 (centre) and NGC\,1281 (right). The extinction in the \IE\ band, $A_{\IE}$ reaches maximum values of 0.4 mag. We transformed the extinction map from $A_{\IE}$ to $A_V$ by using the extinction coefficients calculated as in Sect.~\ref{subsec:integr}, and by assuming a $R_V=3.1$. Then, we transformed $A_V$ in $\Sigma_{\rm dust}$ by using

\begin{equation}
    A_{V} = 0.011 \, \bigg(\frac{Q_{\rm ext}}{2}\bigg)\bigg(\frac{\Sigma_{\rm dust}}{10^3\,\msun\,{\rm kpc}^{-2}}\bigg)\bigg(\frac{\rho}{3{\rm \, g \, cm^{-3}}}\bigg)^{-1}\bigg(\frac{a}{0.1 \, {\rm \mu m}}\bigg)^{-1},
\end{equation}
where $Q_{\rm ext}$ is the extinction coefficient factor, $\rho$ is the dust mass density, and $a$ is the grain radius (e.g. \citealt{kitayama09}). We adopted the standard values given in the equation and we integrated over the disc of dust where $A_V$>0. Assuming that the dust is uniformly distributed over the disc, from the extinction maps we obtain dust masses within the range $\log_{10}(M_{\rm dust, ext}/\msun)\in[5.6, 5.9]$.

We compared these dust masses with those obtained by fitting the integrated SEDs of the targeted galaxies from the FUV to the IR with \texttt{CIGALE}, by including images from \textit{Spitzer}/IRAC and WISE (Sect.~\ref{subsec:suppl}).  
From a first inspection of the IR images, we note that there is no visible signal in the WISE W4 band at 22~$\mu$m, in any galaxy, thus likely indicating a weak dust emission from these galaxies. 
We extracted the flux in the different bands as in Sect.~\ref{subsec:rad_analys}
by using an elliptical region with the SMA equal to their \reff in the $\IE$ band (Table~\ref{tab:radial_data}, \citealt{EROPerseusOverview}), and the position angle and ellipticity coming from the isophote fit. 
For the SED fit, we used the same modules and parameters as listed in Table~\ref{tab:cigale_p}. Fluxes with a ${\rm S/N}<3$ are considered as upper limits and this is the case for all galaxies in the W4 WISE band. The SED fit yields dust masses in the range $\log_{10}(M_{\rm dust, SED}/\msun)\in[5.6, 7.5]$, hence overall consistent or higher than those estimated from the inner disc extinction map. Nevertheless, the dust masses from the SEDs are not well constrained in the fit, i.e. we obtain that the 1$\sigma$ error is larger than the mass value. This might be due to both the lack of FIR data for these galaxies, needed to better constrain the dust emission peak around 100~$\mu$m, or an intrinsically weak concentration of dust. We assume here that the emission of dust in all the analysed ETGs is weak. As an additional test, we performed the SED fitting a second time, without including the \texttt{themis} module. We found that the results of the paper stay unchanged.

\begin{figure*}
\centering
\includegraphics[scale=0.45]{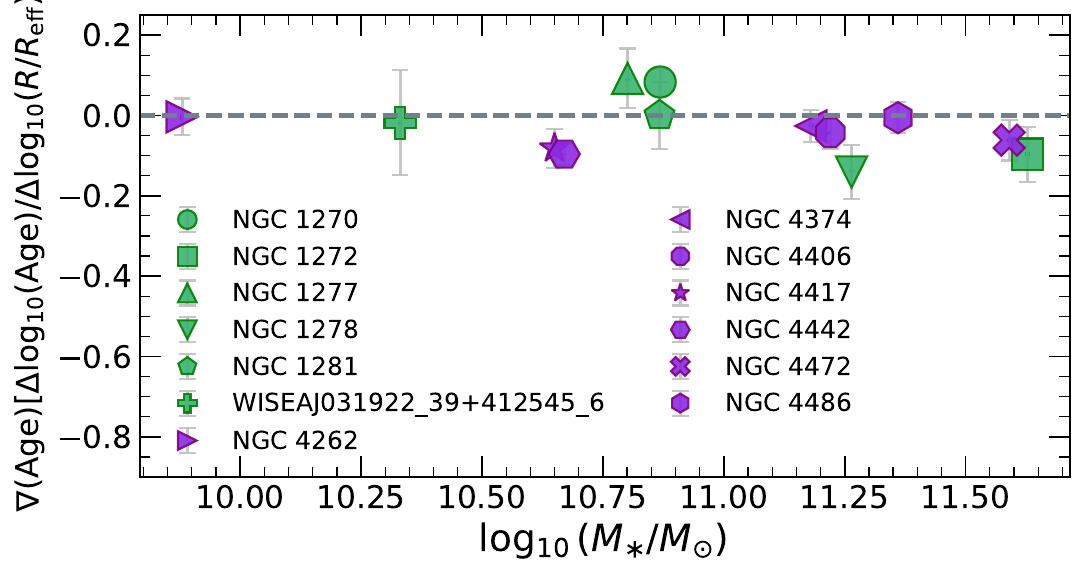}
\hspace{1cm}
\includegraphics[scale=0.45]{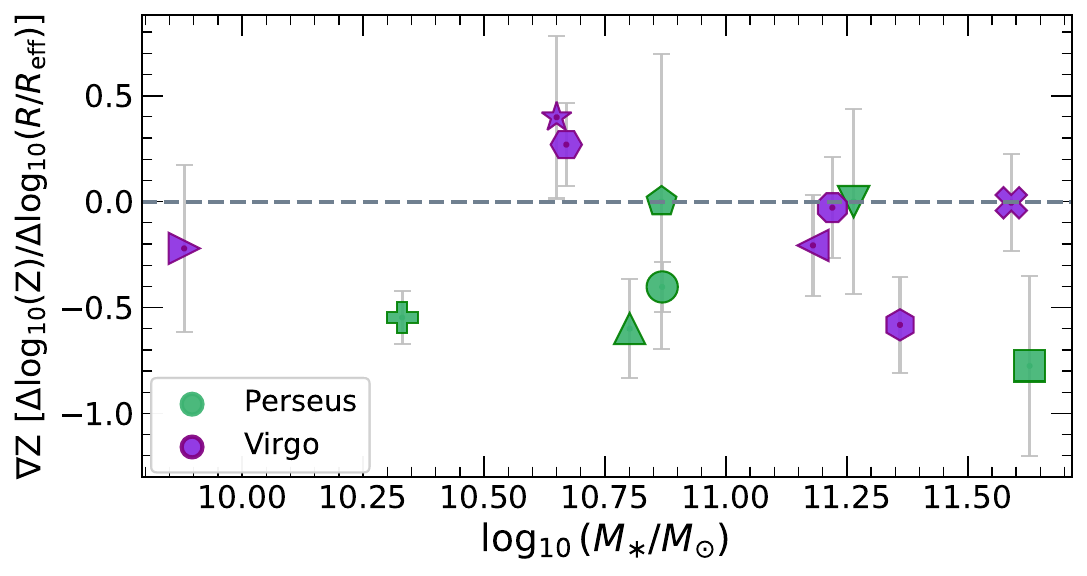}
\caption{Mass-weighted age gradients (\emph{left}) and metallicity gradients (\emph{right}) as a function of galaxy stellar mass, calculated between $R_1$/\reff$=0.1$ and $R_2$/\reff$\in [0.3, 1.6]$. Purple (green) symbols are Virgo (Perseus) galaxies. The horizontal grey dashed line indicates a flat gradient. We note that the gradient of NGC\,1272 is an inner gradient, calculated up to 0.3\reff. See text for more details.}
\label{fig:grads}
\end{figure*}

As explained in Sect.~\ref{subsec:rad_analys}, we masked the central regions of NGC\,1270, 1271, 1272, 1277, 1278, and 1281, for the \texttt{CIGALE} SED fitting (both integrated and radial), due to the presence of AGN in their centre.  We do not mask the central regions for the remaining galaxies in the sample. However, as an additional test, we performed the same SED fitting analysis without masking the central regions of NGC\,1270, 1271, 1272, 1277, 1278, and 1281, and we found that the results of the paper stay unchanged (see Appendix~\ref{subsec:no_mask}).

\section{Results}
\label{sec:res}

\subsection{Gradients of stellar properties}
\label{subsec:gradients_SP}

Figure \ref{fig:res_sed} shows the result of the radial SED fitting. We estimated the properties of the targeted galaxies within the radial sample as a function of their projected SMA, namely the mass-weighted ages, stellar metallicities, temperatures of the UV upturn, and SF timescales. 
As expected (see App.~\ref{app:ind_gal}), the SPs of the galaxies in our sample are old, with mass-weighted ages $\gtrsim 10$ Gyr at all sampled SMAs. Overall, they appear to be more metal-rich (twice solar) in their centre with respect to their outskirt. There are exceptions for NGC\,1278 and NGC\,1281, where the obtained stellar metallicities are flat and half-solar, although for these two galaxies the SMA coverage is rather small. All galaxies need the presence of the UV upturn to explain their FUV emission, with mean temperatures $\langle T_{\rm UV, rad}\rangle = 26\,100\pm 3000$\,K. Finally, the SFH of these galaxies proceeded with an e-folding SF timescale that is $\tau\in[100,1500]$\,Myr, with an average $\langle \tau_{\rm rad}\rangle =770\pm280$\,Myr. For the extended galaxies NGC\,1272 and NGC\,1278 (\reff$>$\,5\,kpc), we note that $\tau$ is on average smaller in the central regions with respect to their outskirts, corresponding to SMA/\reff$\gtrsim0.2$ and $\gtrsim0.4$, respectively. This might likely indicate that the centres of such massive ETGs enriched faster than its peripheries.
To assess the reliability of the stellar parameters derived from the SED fitting, we refer to the mock analysis performed in \citet[their Sect. 5.5 and Appendix E]{martocchia25}, as we used the same method here. In brief, a catalogue of simulated galaxies is generated, where the model fluxes are re-sampled adding gaussian noise. The mock galaxies are then analysed in the same fashion as the real galaxies. All parameters are well recovered, showing strong correlations among the injected input and the recovered output (\citealt{martocchia25}, their Fig.~E.1).

\begin{figure*}
\centering
\includegraphics[scale=0.4]{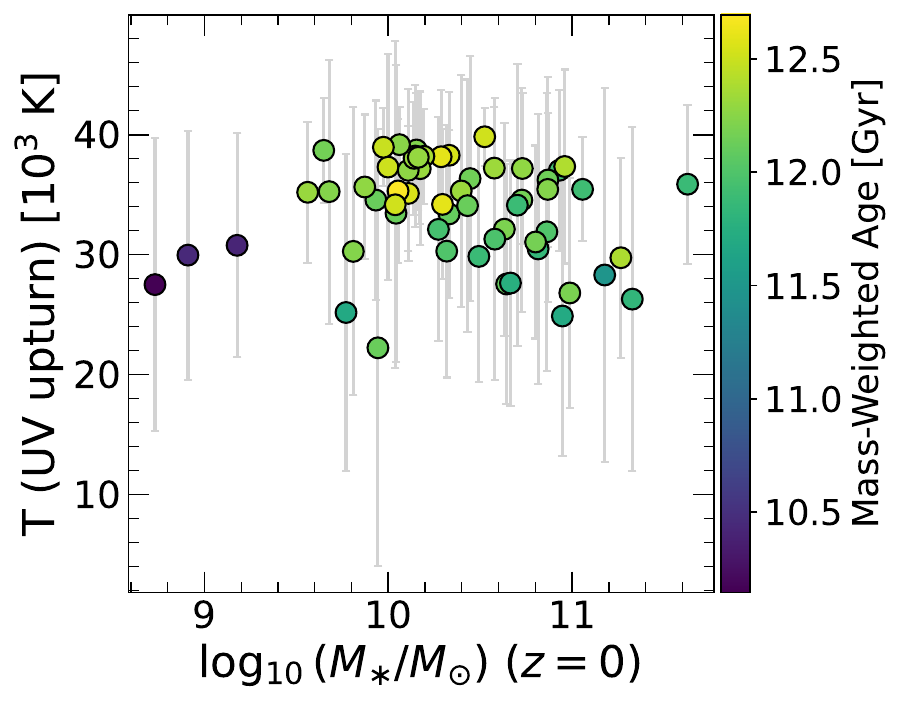}
\hspace{2cm}
\includegraphics[scale=0.4]{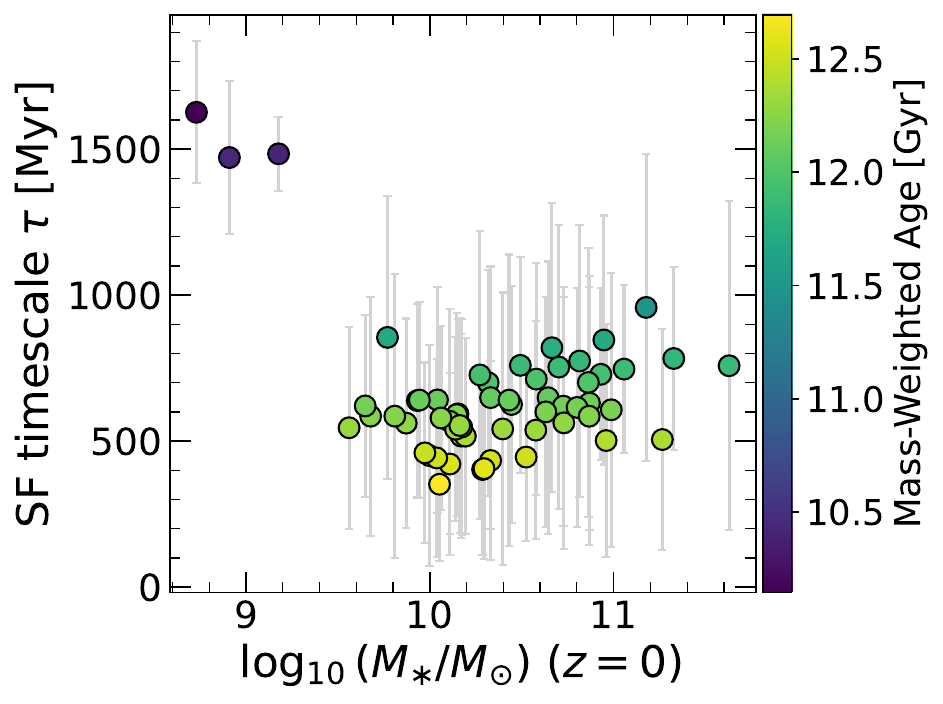}
\caption{Temperature of the UV upturn (\emph{left}), and SF timescale $\tau$ (\emph{right}) as a function of stellar mass (from \citealt{EROPerseusOverview}) for the integrated sample, colour-coded by the mass-weighted ages.}
\label{fig:integr_results_prop}
\end{figure*}
\begin{figure}
\centering
\includegraphics[scale=0.5]{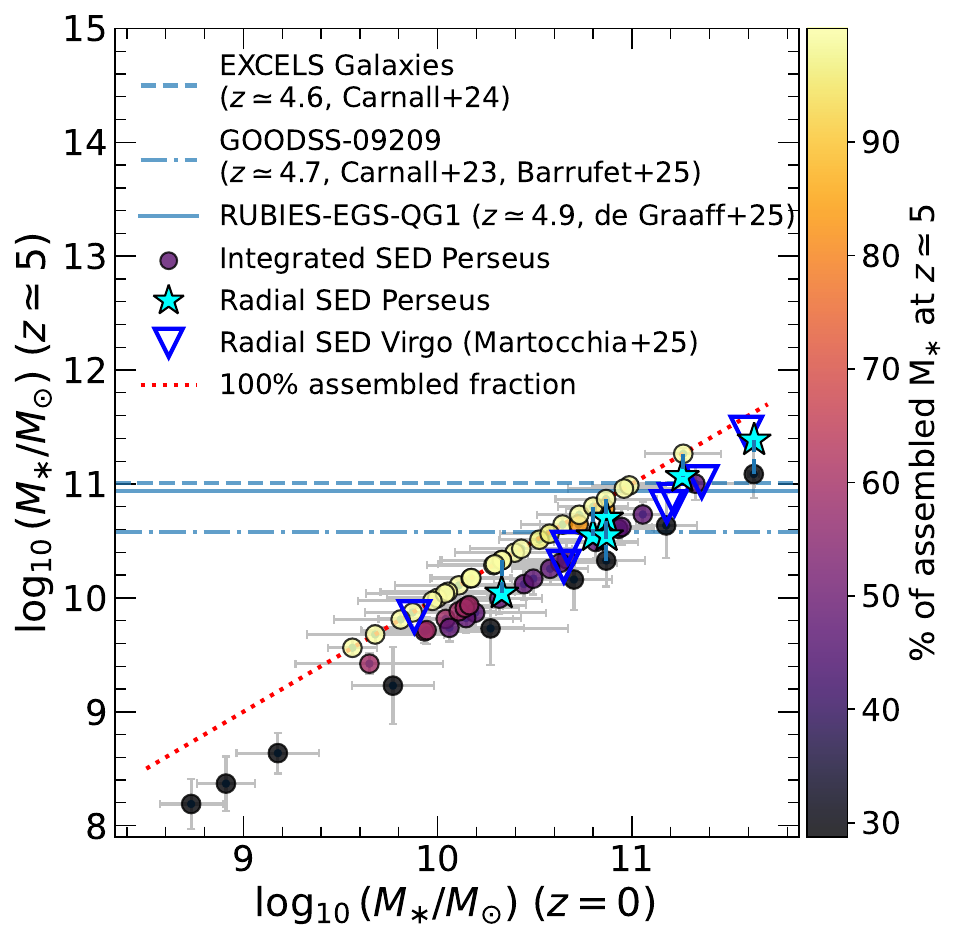}
\caption{Stellar mass at $z=0$ as a function of the estimated stellar mass at $z\simeq5$. Data are colour-coded by the fraction of assembled stellar mass by $z\simeq5$. The Perseus integrated sample is connected by vertical dashed blue lines to the results of the radial analysis (cyan stars). Open blue upside-down triangles indicate the radial results on Virgo massive ETGs by \cite{martocchia25}. 
The red dotted line represents assembled fractions of 100\%, while the solid, dashed, and dot-dashed blue lines indicate the stellar masses found for red quiescent galaxies with JWST at $z>4.6$ by \cite{carnall23, carnall25}, \cite{degraaff25}, and \cite{barrufet25}. Compilation from \cite{antwi-danso23}.}
\label{fig:integr_results}
\end{figure}

Next, we calculated age and metallicity gradients, as they are very useful measurements to discriminate among different formation and evolution scenarios of ETGs (we refer to Sect.~\ref{sec:disc} for more discussion and references). We computed logarithmic gradients for the quantity $X$ as $\Delta\log_{10}(X)/\Delta\log_{10}(R/$\reff) between $R_1$/\reff$=0.1$ and $R_2$/\reff$\in [0.3, 1.6]$, depending on the galaxy, to be able to compare them with Virgo galaxies by \cite{martocchia25}. Figure \ref{fig:grads} shows the mass-weighted age and metallicity gradients as a function of stellar mass for both clusters. We find that gradients in age are typically shallow [$\nabla$(Age)$\gtrsim -0.2$\, dex], comparable between the two clusters. Overall, gradients in $Z$ for Virgo are quite shallow, while for Perseus only 2 out of 6 galaxies have shallow gradients. We indeed found a steep $\nabla Z$ for the extended galaxy NGC\,1272, but we note that we are calculating an inner gradient, between 0.1 and 0.3\,$R/\reff$.
Additionally, we note that within our samples we only have massive ETGs [$\log_{10}(M_{\ast}/\msun)\gtrsim10$]. Shallower gradients for more massive ETGs have been found in a few studies (e.g., \citealt{tortora10, zibetti20, liaocooper23}), however our sample is too small to make a meaningful comparison.
Additionally, we note here that we are calculating gradients for compact galaxies in Perseus, such as NGC\,1270, NGC\,1277 and WISEAJ031922\_30+412545\_6, for which the metallicity gradients are quite steep, i.e. $\nabla Z\lesssim-0.4$\, dex. This is consistent with the spectroscopic works by \cite{trujillo14}, \cite{ferre-mateu17}, and \cite{yilidrim17}. 
Further discussion on stellar gradients is provided in Sects.~\ref{subsec:disc_grads} and \ref{subsec:relic}.

\subsection{The integrated sample}
\label{subsec:res_integr}

Figure \ref{fig:integr_results_prop} shows the $T_{\rm UV}$ (left panel) and $\tau$ (right panel), colour-coded by the mass-weighted ages of the integrated sample. Overall, ETGs need the presence of the UV upturn to explain their FUV emission, with mean temperatures $\langle T_{\rm UV, int}\rangle = 33\,800\pm8500$\,K. Interestingly, the average value is consistent within 1$\sigma$ with the radial sample, though slightly higher. Comparing integrated and radial $T_{\rm UV}$ for the six galaxies in common, the integrated values are generally higher, but agree within uncertainties in four cases. This difference may arise because the radial analysis probes subregions, whereas the integrated study reflects the total galaxy flux. Given the small sample and overall consistency, no firm conclusions can be drawn, and a larger sample is needed to assess whether this trend is general or restricted to a few cases.
We also note that higher mass ETGs tend to have lower $T_{\rm UV}$, but errors are too large to draw strong conclusions. SF timescales for the integrated sample are $\tau\in[250,1000]$\,Myr, except for three low-mass galaxies for which $\tau$ reaches $\simeq1750\,$Myr. The average is $\langle \tau_{\rm int}\rangle=656\pm380\,$Myr.
Hence, these results imply that the massive ETGs in Perseus must have assembled their masses relatively fast. We estimated the fraction of stellar mass that each galaxy in the sample grows at redshift $z\simeq5$, based on the integration of the SFHs that we obtained. Then, we multiplied the stellar mass at $z=0$ from \cite{EROPerseusOverview} by the mass fractions to obtain the stellar mass at $z\simeq5$. Errors at $1\sigma$ on the mass fraction are estimated through the $1\sigma$ errors on the stellar masses from \texttt{CIGALE}. Errors at $1\sigma$ on the stellar masses at $z\simeq5$ are calculated by propagating the stellar mass error from \cite{EROPerseusOverview} and the error on the fractions.
We report the results in Fig.~\ref{fig:integr_results}, where we show the $\Mstellar$ at $z=0$ (from \citealt{EROPerseusOverview}) versus the estimated $\Mstellar$ at $z\simeq5$ for the Perseus integrated sample, colour-coded by the assembled fraction of stellar mass before $z\simeq5$. 
We observe some discrepancy between integrated and radial analysis, with the majority of the cases where the stellar mass at $z\simeq5$ is higher for the integrated study with respect to the radial one. This is likely due to the difference in the aperture radii used in the two studies. Recent works compared resolved and integrated SED fitting techniques and found that the difference in the derived stellar mass is negligible \citep{EP-Abdurrouf}. 
We compare our results with the stellar masses estimated by JWST for red quiescent galaxies at $z\simeq5$ from the compilation by \cite{antwi-danso23}, more specifically from the works by \cite{carnall25}, \cite{degraaff25}, and \cite{barrufet25}.
We note that the majority of galaxies has assembled more than 30--50\% at $z\simeq5$. Around $\simeq$20 galaxies show an extreme accretion close to $\simeq$100\%. Finally, we note that the grown stellar mass in both Perseus and Virgo clusters' ETGs is comparable to that of JWST red quiescent galaxies for the most massive ETGs in the sample, around a mass $\log_{10}(\Mstellar/\msun)\gtrsim 10.8$.

\section{Discussion}
\label{sec:disc}

\subsection{Stellar gradients and the role of mergers}
\label{subsec:disc_grads}
According to the two-phase scenario originally proposed by \cite{oser10}, stars form rapidly in situ, at $z\gtrsim2$, producing a compact galaxy characterized by a steep metallicity gradient ($\nabla Z < -0.35$\,dex; e.g. \citealt{larson74, carlberg84, thomas99, kobayashi04, pipino10}) and a positive age gradient. The latter arises because the galaxy’s deep potential well traps stellar winds, allowing ongoing central star formation and resulting in younger SPs toward the core. Nevertheless, mechanisms such as AGN feedback or morphological quenching \citep{martig09} can suppress central star formation, thus flattening the inner radial profiles (e.g. \citealt{zibetti20}).
In the second phase, ETGs increase their stellar mass through dissipationless mergers, which modify the pre-existing gradients in different ways. Major mergers generally tend to flatten all gradients, as metal-rich stars from the core are redistributed into the outer regions ($\nabla Z \simeq -0.1$\,dex; e.g. \citealt{ogando05, dimatteo09, kobayashi04}). Conversely, minor mergers tend to steepen metallicity gradients, since the accreted satellites are typically more metal-poor than the central regions of the host galaxy (e.g. \citealt{hirschmann15}).

In the literature, it is possible to find a plethora of studies that aimed at constraining the formation of ETGs through the calculation of gradients from both imaging and spectroscopic surveys. 
Informative summaries are found in \cite{goddard17} and in \cite{zibetti20}, and references therein.
We found that the age and metallicity gradients of the galaxies within the sample are overall shallow [$\nabla$(Age/$Z)\gtrsim-0.2$\,dex], except for the compact galaxies, that we will discuss in Sect. \ref{subsec:relic}. However, there are several spectroscopic studies that found steeper Z gradients for ETGs (e.g. \citealt{ martin-navarro18, parikh19, zibetti20} and references therein), with \cite{martin-navarro18} finding that the more massive galaxies show steeper Z gradients with respect to lower mass ones, possibly in contrast with our results. On the contrary, other works found that more massive galaxies show shallower stellar gradients (\citealt{tortora10, zibetti20, liaocooper23}).
First of all, the cited works are based on much larger samples of ETGs spanning different environments. We note that in this work we are analysing ETGs only residing in very dense environments, i.e. galaxy clusters, where gradients have also been found to be shallower \citep{labarbera05}, and have likely been flattened by the higher frequency of major mergers in such harsh environments. As the reported sample of stellar gradients is neither complete nor statistically significant, it would be insightful to enlarge this dataset and compare large samples both in cluster and field environments. 

Additionally, we checked for information on the GC systems of these galaxies. The colour distribution of GCs have indeed been interpreted as the imprint of the two-phase formation scenario of ETGs. The red GCs are more metal-rich and more centrally concentrated: they represent the in situ star formation episode(s); blue or metal-poor GCs, which extend further in the outskirts of galaxies, are thought to be accreted from mergers with smaller satellites (e.g. \citealt{cote98}). While NGC\,1277 have been found to only host a single, red GC population \citep{beasley18}, all the other galaxies\footnote{Except for WISEAJ031922\_39+412545\_6, for which we did not find any information on its GC system, to the best of our knowledge.} in the sample show both a red and blue GC population \citep{alamo-martinez21, harris23}. This demonstrates that all galaxies underwent a rapid phase of in-situ star formation.
Then, the observed shallow gradients might suggest either that major mergers have flattened the gradients during the secular evolution after the in-situ collapse, or AGN feedback had played a role in flattening the gradients \citep{tortora09, tortora10}.

\subsection{Relic galaxies: comparison with recent works}
\label{subsec:relic}

Regarding the compact galaxies in our sample (NGC~1270, NGC~1277, NGC~1281, WISEAJ031922\_39+412545\_6), these are usually considered as good candidate relic galaxies \citep{vandokkum10}\footnote{Among these, NGC~1277 is very likely a relic galaxy \citep{trujillo14, ferre-mateu17, comeron23}.}. Overall, we found that compact galaxies have steeper metallicity gradients compared to more extended ETGs, although this needs to be confirmed with a more significant sample. Because their steep gradients persist to large radii (up to SMA/\reff$\simeq1.6$), these galaxies likely avoided both minor and major mergers, which would otherwise have disrupted their structure and flattened their gradients. For NGC\,1277, we found $\nabla Z=-0.60\pm0.23$\,dex, old SPs and a slightly positive age gradient [$\nabla$(Age)$=0.09\pm0.07$\,dex], a very fast SFH ($\tau<1000$\,Myr), and an assembled mass fraction at $z\simeq5$ close to 100\%. This is consistent with the works by \cite{trujillo14} and \cite{ferre-mateu17}, which measured a $\nabla Z\simeq -0.54$\,dex and $\nabla$(Age)$\simeq 0.02$\,dex.
For NGC\,1270, we also found a steep gradient $\nabla Z=-0.40\pm0.12$\,dex, which is consistent to what was obtained by \cite{yilidrim17}, their Fig. 11. We found instead a slightly subsolar metallicity for NGC\,1281, contrary to what is found by \cite{yildirim16, yilidrim17} who found solar metallicity within the centre of this galaxy (see App.~\ref{app:ind_gal}), as well as steeper gradients. The difference with their work might be due to the small coverage in projected radius available for this work. Also, the discrepancy could be due to the fact that in our models we do not include chemical abundance ratios typical of ETGs, foreseen as a next step in order to improve the upturn SP models. We also observed a very steep gradient for the compact galaxy WISEAJ031922\_39+412545\_6, together with old SPs and large assembled fractions at $z\simeq5$. Hence, WISEAJ031922\_39+412545\_6 might be considered a good candidate relic galaxy. Similarly, many other galaxies within the Perseus cluster might show a high degree of relicness \citep{ferre-mateu17}. As defined in \cite{spiniello24}, these are ETGs that assembled more than 75\% of their stellar masses by $z=2$. However, these considerations need to be confirmed through more in-depth structural and kinematical analysis.

\subsection{Comparison with high-z red quiescent galaxies from JWST}
\label{subsec:comp_jwst}

Within our sample, we are looking for ETGs in the local Universe that might represent the descendants of the most massive red and quiescent galaxies observed by JWST at $z>4.6$. It is important to consider that local ETGs have larger sizes than red galaxies observed at high redshfits (e.g. \citealt{daddi05}), therefore a one-to-one comparison could only be performed with the relic galaxies. However,
determining the number of nearby ETGs with formation histories analogous to massive high-$z$ red, quiescent galaxies is crucial, as their resolved SPs and SFHs might provide key constraints on cosmological and galaxy formation models. The reconstructed SFHs of the ETGs in this work show that the most massive galaxies in the sample [with $\log_{10}(M_{\ast}/\msun)>10.8$] might have had comparable stellar masses, at $z\simeq5$, than the observed massive red quiescent galaxies observed by JWST at similar redshifts (Fig.~\ref{fig:integr_results}). We found 19 ETGs within the sample (including Virgo galaxies) that had $\log_{10}(M_{\ast}/\msun)>10.6$ at $z\simeq5$, comparable to the galaxies by \cite{carnall23,carnall25}. Additionally, we found that only 5 local ETGs had extreme stellar masses $\log_{10}(M_{\ast}/\msun)>11$ at $z\simeq5$ comparable to the most extreme case found by JWST \citep{degraaff25}. We suggest then that we could look for the progeny of the quiescent JWST high-mass galaxies among the most massive ETGs, which is consistent with what was already predicted by \cite{thomas05}.

\section{Summary and Conclusions}
\label{sec:concl}
We combined the exceptional image quality and depth of the ERO Perseus with FUV and NUV observations from GALEX and AstroSat/UVIT, as well as with $ugriz\Ha$ data from MegaCam at the CFHT, to deliver FUV-to-NIR magnitudes of the 87 brightest galaxies within the Perseus cluster (Table~\ref{tab:fuv_nuv_integr}).
We compared the integrated colours of Perseus in ${\rm FUV-NUV}$ and $\textrm{FUV}-\IE$ with those of the Virgo cluster galaxies in Fig.~\ref{fig:fuv_nuv_integr}, showing that Perseus and Virgo ETGs have similar stellar mass-colour diagrams. On the contrary, we observe different ${\rm FUV-NUV}$ and $\textrm{FUV}-\IE$ colours for the LTGs, with Perseus LTGs being much redder and less numerous (at similar stellar masses) than Virgo's LTGs. This suggests an efficient ongoing quenching process due to ram-pressure stripping \citep{boselli14}.

We reconstructed the SFH of the ETGs in the sample through the SED fitting code \texttt{CIGALE}. For a subsample of the galaxies with stellar masses $\log_{10}(\Mstellar/\msun)>10.3$, we analysed their radial stellar properties. We combined a high angular resolution, high-sensitivity multiwavelength dataset (FUV-to-NIR) with state-of-the-art SP models, which include a flexible UV-upturn modelling as a function of fuel and temperature (\citealt{mt00, lecras16, martocchia25}).
We studied radial profiles that span from the galaxy nuclei (SMA/$\reff \simeq 0.1$) up to 1.5\,\reff for the majority of these ETGs. We showed that, despite the lack of ongoing SF, the ETGs in the sample have non-negligible UV emission (Fig.~\ref{fig:rad_UV}) which is comparable to that of Virgo most massive ETGs. This is most likely due to old, low-mass stars in late evolutionary stages, i.e. the UV upturn. 
For the majority of ETGs within the sample, we found indeed that the SEDs of Perseus ETGs are well described by an UV upturn component. This might suggest that the UV upturn is an intrinsic property of ETGs as it was proposed by \cite{boissier18}. 
The average temperatures we found for the radial sample are $\langle T_{\rm UV,rad}\rangle =26\,100\pm3000$\,K, slightly lower, albeit comparable, with what we found for Virgo ETGs in \cite{martocchia25}. For the integrated sample we found $\langle T_{\rm UV, int}\rangle= 33\,800\pm8500$\,K. 
We also found that the SF timescales derived for all ETGs are relatively small. For the radial sample, we estimated timescales $\tau\in[100,1500]$\,Myr, while for the full sample $\tau\in[250,1700]$\,Myr. Additionally, we found shallow age and metallicity gradients, except for the compact galaxies within the sample. We speculate that the shallow gradients could be driven by AGN feedback or major mergers, while a steep behavior in metallicity is consistent with the inside-out growth scenario, where the densest central regions collapse and form stars on very short timescales (e.g. \citealt{pipino10}). 
For the compact galaxies, we then found results which are consistent with the view that such galaxies might be the direct descendants (relics) of the red compact galaxies found at high-$z$, remaining untouched by mergers. 
Through the retrieved SFH with \texttt{CIGALE}, we estimated the fraction of formed/assembled stellar mass at $z\simeq5$ for all the ETGs in the sample (see Fig.~\ref{fig:integr_results}). We show that for present-day stellar masses of $\log_{10}(\Mstellar/\msun)\gtrsim 10.8$, these galaxies had masses at $z\simeq5$ comparable to those reported by recent JWST studies of red quiescent galaxies at similar redshifts (\citealt{antwi-danso23} and references therein). Our results, together with those regarding the most massive ETGs in Virgo, might suggest that massive local ETGs inhabiting massive clusters [total mass $\log_{10}(M/\msun)\gtrsim 10^{14}\msun$] enriched fast and in a similar way to red massive galaxies at high redshift, likely representing their progeny.

Nevertheless, the reported works for Virgo and Perseus, which are representative of massive local clusters, serve as benchmark projects that will pave the way to similar studies of a much larger sample of ETGs in diverse environments, from the richest galaxy clusters down to cluster groups, filaments, and voids. This study will be complementary to several other techniques already being studied by the \Euclid\ community, such as in \cite{EP-Kovacic}, \cite{EP-Abdurrouf}, and \cite{EP-Nersesian} to estimate the physical properties of galaxies mapped with an unprecedented sky coverage (14\,000\,deg$^2$), angular resolution, and image quality in terms of sensitivity that \Euclid\ will soon provide. This will also enable SP analyses in combination with the detection of LSB features down to 30 mag\,arcsec$^{-2}$, as well as their GC population (e.g. \citealt{saifollahi25, Larsen25}) to reconstruct the assembly histories of a wide sample of ETGs.  
These datasets will be coupled with other upcoming surveys such as the \textit{Rubin}/LSST, as well as FUV and NUV observations: the GALEX All-sky Imaging Survey offers a large sky coverage, more than $\simeq$22\,000\,deg$^2$ \citep{bianchi17}, and with a limiting magnitude of $21$--22; the GALEX Medium-depth Imaging Survey is also available, reaching deeper limiting magnitudes of $\simeq$23 \citep{bianchi14}. 
In addition, observations in the FUV with the AstroSat are ongoing 
with more than 1700 targets observed so far \citep{bordoloi24}. 
With AstroSat/UVIT we have already obtained approved programmes of several ks to expand the Perseus UV footprint (PI S. Martocchia) as well as for the Fornax ERO field (PI T. Woods). The detection of FUV emission in ETGs, coupled with the lack of gas emission, and consequently of residual SF, highlights the ubiquity of UV upturn stars, whose origin remains a subject of debate. This demonstrates the need for future sensitive FUV observatories, such as the Cosmological Advanced Survey Telescope for Optical and Ultraviolet Research (CASTOR; \citealt{cote25}), the UltraViolet EXplorer (UVEX; \citealt{kulkarni21,fucik24}), and NASA’s upcoming flagship mission, the Habitable Worlds Observatory, which will be key for studies of SPs in galaxies, including the origin of UV upturn stars (\citealt{smercina25}). In parallel, next-generation FIR facilities with high angular resolution, such as the proposed PRobe far-Infrared Mission for Astronomy (PRIMA; \citealt{ciesla25}), will also be needed, to spatially resolve dust emission in galaxies and extend the wavelength coverage of such high-resolution studies to the FIR.

\section{Data Availability}
Tables~\ref{tab:fuv_nuv_integr}, \ref{tab:optical_integr_galex}, and \ref{tab:optical_integr_uvit} are available only in electronic form at the CDS via anonymous ftp to \url{cdsarc.u-strasbg.fr} (130.79.128.5) or via \url{http://cdsweb.u-strasbg.fr/cgi-bin/qcat?J/A+A/}.

\begin{acknowledgements}
We thank the anonymous referee for a constructive report. We also thank C. Carvajal-Bohorquez and O. Ilbert for useful discussions. M.Boquien acknowledges support from the ANID BASAL project FB210003. This work was supported by the French government through the France 2030 investment plan managed by the National Research Agency (ANR), as part of the Initiative of Excellence of Universit{\'e} C{\^o}te d'Azur under reference number ANR-15-IDEX-01.
  \AckERO \AckEC 
\end{acknowledgements}

\bibliography{biblio}

\appendix 
\onecolumn

\section{Continuum-subtracted H\texorpdfstring{$\alpha$}{alpha} image of NGC\texorpdfstring{\,}{}1275}
\begin{figure*}[h!]
\centering
\includegraphics[scale=0.72]{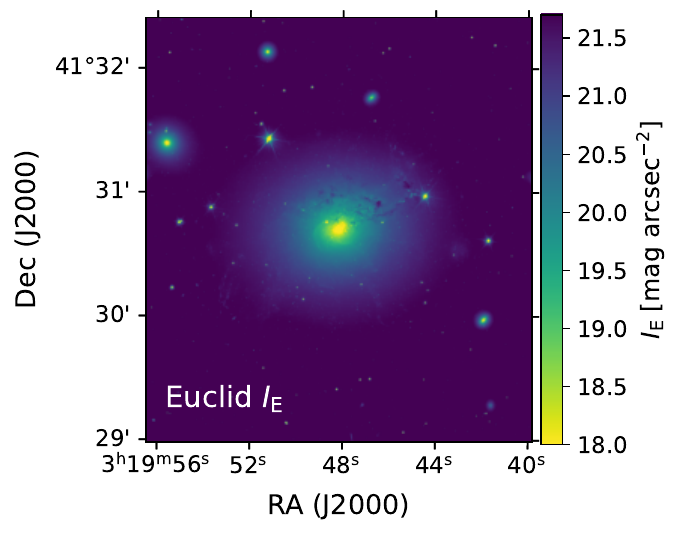}
\includegraphics[scale=0.72]{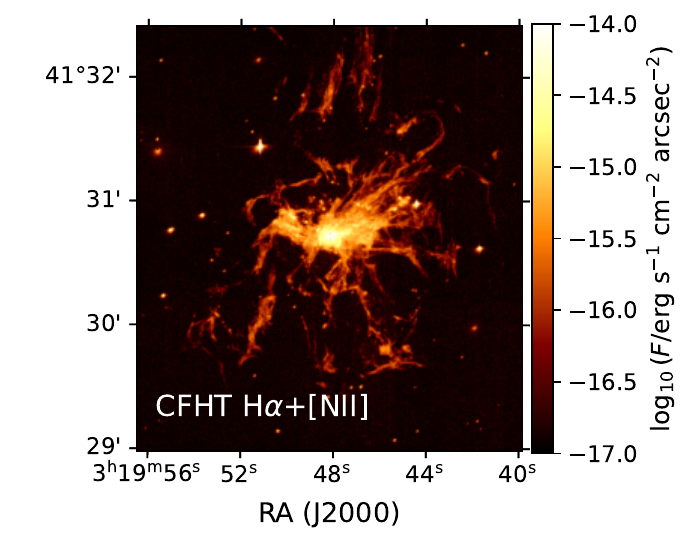}
\caption{\textit{Euclid \IE} (\emph{left}) and continuum-subtracted \Ha (\emph{right}) image of the central galaxy of Perseus, NGC\,1275.} 
\label{fig:ha}
\end{figure*}

Figure~\ref{fig:ha} shows the central galaxy of the Perseus cluster, NGC\,1275, observed in the $\IE$ band in the left panel, while the right panel reports the continuum-subtracted \Ha image. The exceptional details of the $\IE$ image reveals the presence of dust filaments located north-west of the galactic centre, as well as faint stellar structures such as tidal tails located east of the centre of NGC\,1275. 
The continuum-subtracted \Ha CFHT image, instead, reveals the well-known filaments characteristic of the cooling flows of this galaxy (e.g. \citealt{conselice01}). 

\section{Integrated UV magnitudes of Perseus galaxies within \Euclid}
\label{app:table_uv_integr}

Tables~\ref{tab:fuv_nuv_integr}, \ref{tab:optical_integr_galex}, and \ref{tab:optical_integr_uvit} report the estimated magnitudes from the FUV to the NIR on the galaxy sample in the Perseus ERO field of view with ${\rm S/N}>3$ in both FUV and NUV bands. Table~\ref{tab:fuv_nuv_integr} reports the FUV and NUV magnitudes along with their 1$\sigma$ errors in both the GALEX and UVIT sample, while Table~\ref{tab:optical_integr_galex} (\ref{tab:optical_integr_uvit}) reports the estimated optical and NIR magnitudes in the GALEX (UVIT) sample. Table~\ref{tab:fuv_nuv_integr} and \ref{tab:optical_integr_galex} consist of 87 rows, while Table~\ref{tab:optical_integr_uvit} consists of 28 rows.

\begin{landscape}
\begin{table}
\caption{FUV and NUV photometry of the integrated galaxy sample.}
\begin{tabular}{rlrrrrrrrrrrrr}
\hline\hline
ID & Galaxy & RA & Dec & $R_{\sfont{G}}$ & $\textrm{FUV}_{\sfont{G}}$ & eFUV$_{\sfont{G}}$ & NUV$_{\sfont{G}}$ & eNUV$_{\sfont{G}}$ & $R_{\sfont{U}}$ & FUV$_{\sfont{U}}$ & eFUV$_{\sfont{U}}$ & NUV$_{\sfont{U}}$ & eNUV$_{\sfont{U}}$ \\
   &        & (deg)& (deg)& ($\arcsecond$)  & (mag)   & (mag)    & (mag)   & (mag)  & ($\arcsecond$)  & (mag)   & (mag)   & (mag) & (mag) \\
(1) & (2) & (3) & (4) & (5) & (6) & (7) & (8) & (9) & (10) & (11) & (12) & (13) & (14) \\
\hline
\small
  1 & MCG+07-07-070 & 50.0919121 & 41.6409150 & 14.86 & 16.49 & 0.05 & 16.28 & 0.06 & 16.52 & 16.34 & 0.06 & 16.02 & 0.02\\
  2 & NGC\,1275 & 49.9507129 & 41.5116267 & 86.22 & 14.72 & 0.08 & 14.20 & 0.11 & 79.73 & 14.45 & 0.07 & 13.93 & 0.04\\
  3 & NGC\,1272 & 49.8387080 & 41.4906450 & 9.38 & 19.44 & 0.08 & 18.87 & 0.12 & 5.62 & 19.66 & 0.09 & 19.26 & 0.05\\
  $...$ & $...$ & $...$ & $...$ & $...$ & $...$ & $...$ & $...$ & $...$ & $...$ & $...$ & $...$ & $...$ & $...$\\
\hline
\label{tab:fuv_nuv_integr}
\end{tabular}
\tablefoot{
Columns give the following information: (1) ID, (2) galaxy name from \cite{EROPerseusOverview}, (3) RA (J2000), (4) Dec (J2000) from \cite{EROPerseusOverview}, (5) radius in arcsec, used for the calculation of the magnitudes on the GALEX images, (6), (7), (8), (9) GALEX FUV, NUV magnitudes with 1$\sigma$ errors; (10) radius in arcsec, used for the calculation of the magnitudes on the UVIT images, (11), (12), (13), (14) AstroSat/UVIT FUV, NUV magnitudes with 1$\sigma$ errors.
}
\end{table}

\begin{table}
\caption{Optical and NIR photometry of the integrated GALEX galaxy sample.}
\begin{tabular}{lrrrrrrrrrrrrrrrrrr}
\hline\hline
ID & $u_{\sfont{G}}$ & e$u_{\sfont{G}}$ & $g_{\sfont{G}}$ & e$g_{\sfont{G}}$ & $r_{\sfont{G}}$ & e$r_{\sfont{G}}$ & $i_{\sfont{G}}$ & e$i_{\sfont{G}}$ & $z_{\sfont{G}}$ & e$z_{\sfont{G}}$ & $I_{\sfont{E,G}}$ & e$I_{\sfont{E,G}}$ & $Y_{\sfont{E,G}}$ & e$Y_{\sfont{E,G}}$ & $J_{\sfont{E,G}}$ & e$J_{\sfont{E,G}}$ & $H_{\sfont{E,G}}$ & e$H_{\sfont{E,G}}$\\
 & (mag)  & (mag)   & (mag)    & (mag)   & (mag)  & (mag)  & (mag)   & (mag)   & (mag) & (mag) & (mag)& (mag)& (mag)& (mag)& (mag)& (mag)& (mag) & (mag)\\
(1) & (2) & (3) & (4) & (5) & (6) & (7) & (8) & (9) & (10) & (11) & (12) & (13) & (14) & (15) & (16) & (17) & (18) & (19)\\
\hline
\small
  1 & 15.399 & 0.113 & 14.938 & 0.156 & 14.563 & 0.189 & 14.631 & 0.270 & 14.636 & 0.336 & 14.757 & 0.253 & 14.584 & 0.341 & 14.565 & 0.230 & 14.515 & 0.220\\
  2 & 12.824 & 0.119 & 11.631 & 0.092 & 10.774 & 0.045 & 10.580 & 0.082 & 10.520 & 0.064 & 10.784 & 0.066 & 10.173 & 0.045 & 10.068 & 0.041 & 9.918 & 0.040\\
  3 & 15.661 & 0.057 & 13.808 & 0.029 & 12.780 & 0.018 & 12.435 & 0.015 & 12.186 & 0.011 & 12.761 & 0.014 & 12.008 & 0.011 & 11.865 & 0.010 & 11.710 & 0.011\\
  $...$ & $...$ & $...$ & $...$ & $...$ & $...$ & $...$ & $...$ & $...$ & $...$ & $...$ & $...$ & $...$ & $...$ & $...$ & $...$ & $...$ & $...$\\
\hline
\label{tab:optical_integr_galex}
\end{tabular}
\tablefoot{Columns give the following information: (1) ID, (2)-(19) CFHT $ugriz$, $\IE, \YE, \JE, \HE$ magnitudes with 1$\sigma$ errors for the GALEX apertures.}
\end{table}

\begin{table}
\caption{Optical and NIR photometry of the integrated UVIT galaxy sample.}
\begin{tabular}{lrrrrrrrrrrrrrrrrrr}
\hline\hline
ID & $u_{\sfont{U}}$ & e$u_{\sfont{U}}$ & $g_{\sfont{U}}$ & e$g_{\sfont{U}}$ & $r_{\sfont{U}}$ & e$r_{\sfont{U}}$ & $i_{\sfont{U}}$ & e$i_{\sfont{U}}$ & $z_{\sfont{U}}$ & e$z_{\sfont{U}}$ & $I_{\sfont{E,U}}$ & e$I_{\sfont{E,U}}$ & $Y_{\sfont{E,U}}$ & e$Y_{\sfont{E,U}}$ & $J_{\sfont{E,U}}$ & e$J_{\sfont{E,U}}$ & $H_{\sfont{E,U}}$ & e$H_{\sfont{E,U}}$\\
 & (mag)  & (mag)   & (mag)    & (mag)   & (mag)  & (mag)  & (mag)   & (mag)   & (mag) & (mag) & (mag)& (mag)& (mag)& (mag)& (mag)& (mag)& (mag) & (mag)\\
(1) & (2) & (3) & (4) & (5) & (6) & (7) & (8) & (9) & (10) & (11) & (12) & (13) & (14) & (15) & (16) & (17) & (18) & (19)\\\hline
\hline
\small
  1 & 15.330 & 0.108 & 14.771 & 0.178 & 14.518 & 0.191 & 14.757 & 0.247 & 14.668 & 0.341 & 14.688 & 0.252 & 14.519 & 0.335 & 14.530 & 0.306 & 14.475 & 0.278\\
  2 & 12.853 & 0.112 & 11.568 & 0.078 & 10.852 & 0.047 & 10.844 & 0.093 & 10.678 & 0.063 & 10.837 & 0.066 & 10.230 & 0.048 & 10.123 & 0.041 & 9.969 & 0.034\\
  3 & 16.158 & 0.057 & 14.190 & 0.010 & 13.267 & 0.007 & 13.098 & 0.010 & 12.746 & 0.006 & 13.224 & 0.008 & 12.460 & 0.005 & 12.312 & 0.005 & 12.155 & 0.004\\
  $...$ & $...$ & $...$ & $...$ & $...$ & $...$ & $...$ & $...$ & $...$ & $...$ & $...$ & $...$ & $...$ & $...$ & $...$ & $...$ & $...$ & $...$\\
\hline
\label{tab:optical_integr_uvit}
\end{tabular}
\tablefoot{Same as in Table~\ref{tab:optical_integr_galex} but for the UVIT apertures.}
\end{table}

\end{landscape}

\section{Comparison between GALEX and UVIT magnitudes.}
\label{app:comp_gal_uv}
\begin{figure*}[h!]
\centering
\includegraphics[scale=0.6]{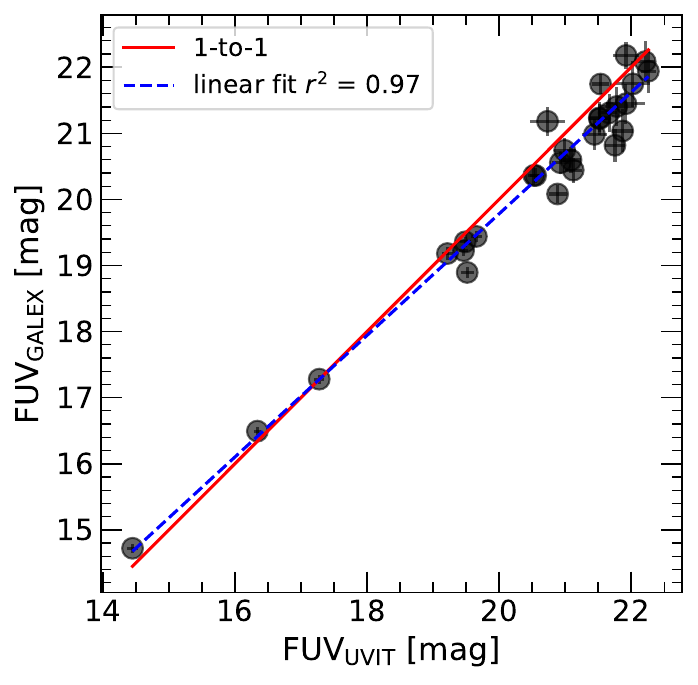}
\includegraphics[scale=0.6]{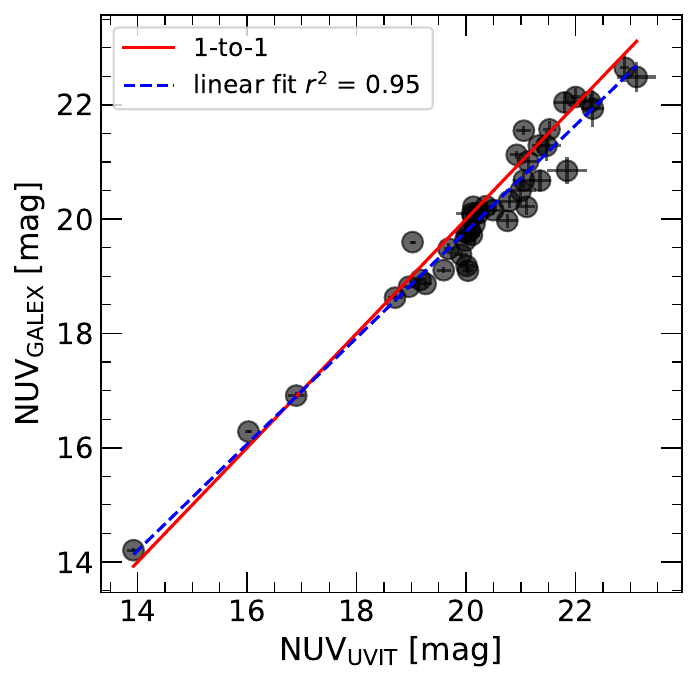}
\caption{Comparison between the UVIT and GALEX FUV (\emph{left}) and NUV (\emph{right}) magnitudes for galaxies with a ${\rm S/N}>3$ in the FUV and NUV bands, respectively. The solid red line indicates the one-to-one relation while the dashed blue line represents a linear fit to the data. The squared Pearson correlation coefficient $r^2$ is reported.}
\label{fig:comp_UV}
\end{figure*}

The left (right) panel of Fig.~\ref{fig:comp_UV} reports the comparison among the FUV (NUV) magnitudes from GALEX and UVIT. We show a 1-to-1 relation as well as a linear fit. Figure~\ref{fig:comp_UV} shows that the FUV and NUV magnitudes from GALEX and UVIT are comparable, as the correlation coefficient $r^2$ is $>0.95$.

\section{Notes on individual galaxies}
\label{app:ind_gal}
In this Appendix, we provide complementary information from the recent literature on the individual galaxies for which we performed a spatially-resolved SP analysis.

\textit{NGC\, 1270.}
Classified as an E3 galaxy by \cite{meusinger20}, it is a compact galaxy (\reff $\simeq 2.6$\,kpc, \citealt{EROPerseusOverview}) that has been found to be a fast rotator by \cite{raskutti14}. It is part of the Hobby-Eberly Massive Galaxy Survey (HEMGS), which targeted high velocity dispersion galaxies which might host ``over-massive'' super massive black holes (SMBH). \cite{ferre-mateu15} found NGC\,1270 to be an outlier in the $M_{\rm BH}$-$\sigma$ relation, together with NGC\,1277 and NGC\,1281.
Spectroscopic integral-field unit SP analysis also found a shallow gradient up to 2\reff by \cite{greene12}. 

\textit{NGC\, 1272.} This galaxy is the second brightest and most massive ETG of the Perseus cluster. It was found to host a SMBH in its centre by \cite{Saglia24}, with a powerful radio jet \citep{gendron-marsolais21}. It is kinematically classified as a slow rotator. 

\textit{NGC\, 1277.} NGC\,1277 was extensively studied in the literature being the first proposed candidate as a relic galaxy, now considered the perfect example of this pristine class of galaxies \citep{trujillo14}. NGC\,1277 is indeed compact (\reff < 2\, kpc) and it has been found to lack evidence of ongoing star-formation \citep{salvador-rusinol22}. Its SPs are very old as well as iron- and $\alpha$-rich \citep{ferre-mateu17}. It has been found to host a single red GC population \citep{beasley18} and to have no signs of tidal features by \cite{trujillo14}, indicative of its nature untouched by mergers. 
Additionally, it has been found to have a bottom-heavy IMF \citep{martin-navarro15} and to be dark matter deficient up to 5\reff, which has been interpreted as the reason why NGC\, 1277 has not accreted a halo. The lack of dark matter would inhibit dynamical friction and thus the accretion of smaller dwarf satellites \citep{comeron23}. 
\cite{emsellem13} and \cite{comeron23} also classified it as a fast rotator based on its stellar kinematic. 

\textit{NGC\, 1278.} NGC\,1278 is an elliptical (E3) galaxy (\reff $\simeq5.6$\,kpc, \citealt{EROPerseusOverview}) that was kinematically studied by \cite{comeron23} in order to compare it to its neighbour NGC\,1277 (see Fig. \ref{fig:sb_vis}). \cite{comeron23} found that NGC\,1278 is slow rotator and it has a bottom-lighter IMF with respect to NGC\,1277. Additionally, the two galaxies differ in their dark matter content, with NGC\,1278 having a non-negligible dark matter fraction. 

\textit{NGC\, 1281.} NGC\,1281 is also a compact galaxy, possible relic candidate. It is a fast rotator, according to \cite{yildirim16}, and lie above the $M_{\rm BH}-\sigma$ relation, together with NGC\,1277 and NGC\,1270 \citep{ferre-mateu15}. Additionally, \cite{yildirim16} found that NGC\,1281 is dark matter rich, three times higher with respect to ETGs in the ATLAS$^{3D}$ sample from \cite{emsellem11}. \cite{ferre-mateu15} also found solar metallicity in the centre of the galaxy which decreases to subsolar at 3\reff. 

\textit{WISEAJ031922\_39+412545\_6.}
S0 compact galaxy classified as passive (no AGN) by \cite{meusinger20}. To the best of our knowledge, we are not aware on any specific studies on this galaxy. We did not find information neither on its stellar kinematic nor on its SPs.

\section{Table of adopted \texttt{CIGALE} parameters}
\label{app:table_cigale}
Table~\ref{tab:cigale_p} reports on the adopted \texttt{CIGALE} modules and parameters used to fit the galaxies' SEDs. 

\begin{table*}[h!]
\centering
\caption{CIGALE adopted parameters and properties to fit the SEDs of the target ETGs.}
\begin{tabular}{c l c l}
\hline
Parameter & Value & Units & Description \\
\hline
\multicolumn{4}{c}{SFH: \texttt{sfhdelayed} module} \\ 
$t$ & $0$--13\,400  & Myr & SFH range\\
$\tau$ & 100, 300, 500, 800, 1000,  & Myr & e-folding timescale of SF\\
& 1500, 2000, 2500, 3000, 4000, 5000 & & \\
\hline
\multicolumn{4}{c}{SSPs: \texttt{uvupturn} module} \\
$Z$ & 0.01, 0.02, 0.04 & & Metallicity\\
$T$ & 0 (no Upturn), 25\,000, 35\,000, 40\,000 & K & Temperature of the upturn component\\
$f$ & 0 & & Fuel parameter. 0: low fuel, 1: high fuel. \\
\hline
\multicolumn{4}{c}{Dust Attenuation: \texttt{dustatt\_modified\_starburst} module} \\
$E(B-V)_{\rm young}$ & 0.0, 0.005, 0.01, 0.02, 0.05, 0.1, 0.3, & & Stellar colour excess of the young populations.\\
 & 0.4, 0.5, 0.7, 0.8, 0.9, 1.0 & mag &  \\
$E(B-V)_{\rm old}$ & 0, 0.0025, 0.005, 0.01, 0.025, 0.05, & & Stellar colour excess of the old populations.\\
  & 0.15, 0.2, 0.25, 0.35, 0.4, 0.45, 0.5 & mag & \\
$E(B-V)$ factor &  0.5 & mag & Conversion factor between young and old.\\
uv\_bump\_amplitude &  3 &  & Amplitude of the UV bump\\
powerlaw\_slope & $-0.25$  & & Slope $\delta$ of the attenuation powerlaw\\
\hline
\multicolumn{4}{c}{Dust emission: \texttt{themis} module} \\
$q_{\rm hac}$ & 0.02, 0.06, 0.14, 0.17, & & Mass fraction of hydrocarbons solids (HAC)\\
         & 0.2, 0.24, 0.32, 0.4 &  & \\
$U_{\rm min}$ & 0.1, 0.3, 0.5, 0.8, 1.2, 1.5, 2, 2.5, & & Minimum radiation field\\
 & 3, 3.5, 6, 10, 17, 30, 50, 80 & Habing &  \\
$\alpha$ & 1.0, 1.5, 2.0, 2.5 & & powerlaw slope\\ 
$\gamma$ & 0, 0.001, 0.002, 0.004, 0.008, & & Fraction illuminated from $U_{\rm min}$ to $U_{\rm max}$\\
 & 0.016, 0.031, 0.063, 0.13, 0.25, 0.5 & & \\
\hline
\multicolumn{4}{c}{Properties} \\
sfh.sfr & from CFHT-\Ha images & $M_{\sun}/{\rm yr}$ & 5$\sigma$ upper limit on SFR.\\
\hline
\end{tabular}
\label{tab:cigale_p}
\end{table*}

\section{SED fitting results of radially analysed galaxies}
\label{app:sed_fitting_radial_res}
\begin{figure}[h!]
\centering
\includegraphics[scale=0.55]{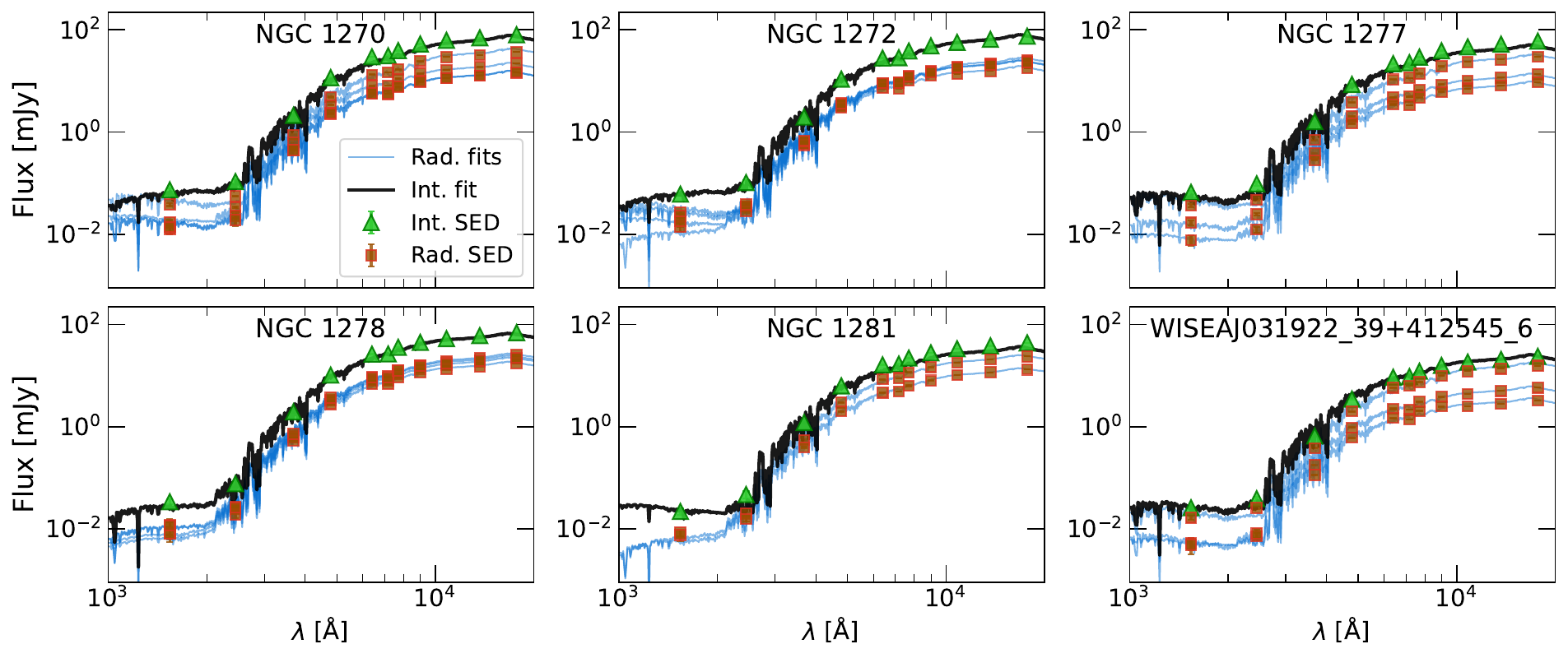}
\caption{Radial (integrated) SEDs for the six galaxies within the radial sample. Brown squares (green triangles) represent the observed fluxes for the radial (integrated) fit, while solid blue (black) lines indicate the best-fit models for the radial (integrated) fits.}
\label{fig:res_sed_radial}
\end{figure}

Figure~\ref{fig:res_sed_radial} shows the radial and integrated SED fitting results for the six galaxies in the radial sample.

\section{SED fitting quality tests}
\label{app:sed_fitting_qual}

\subsection{SED fitting with no UV Upturn and contribution from young stellar populations}
\label{subsec:noup_sedfit}
\begin{figure}[h!]
\centering
\includegraphics[scale=0.52]{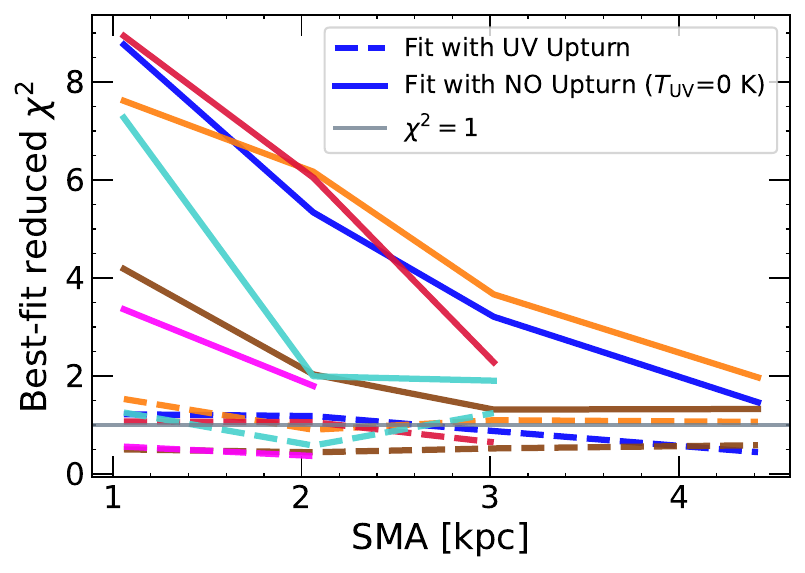}
\hspace{1cm}
\includegraphics[scale=0.52]{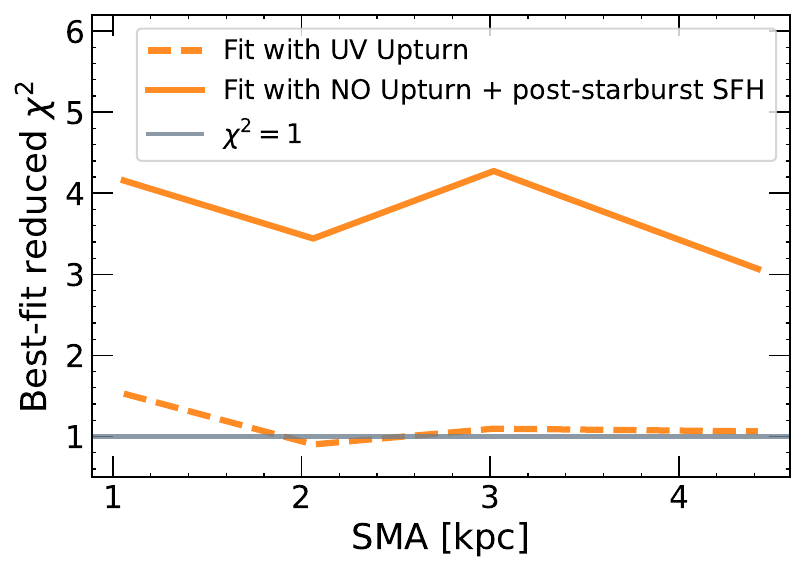}
\caption{\emph{Left}: Best-fit reduced $\chi^2$ as a function of SMA for the galaxies analysed radially. Colours are as in Fig.\ref{fig:rad_UV}. Solid (dashed) lines indicate fits performed without (with) an UV upturn component. \emph{Right}: as in the left panel, but only for NGC\,1272. In this case, the solid line indicates a no Upturn model with the addition of a post-starburst component in the SFH.}
\label{fig:chi_square}
\end{figure}

To test whether the UV emission could be reproduced by a young stellar component, we first checked the quality of the fit when setting $T_{\rm UV}=0$\,K (no UV upturn component), both for the integrated and the radial sample. Overall, we observe that the quality of the fit worsens compared to when upturn models are employed (Fig.~\ref{fig:chi_square}, left). 
Then, in this case where $T_{\rm UV}=0$\,K, we calculated the fraction in mass of stars younger than 2~Gyr through \texttt{CIGALE}. \cite{salvador-rusinol22} indeed found an anti-correlation between the stellar mass and the fraction of stars younger than 2 Gyr. We found that the two lowest-mass galaxies in the sample [$\log_{10}(\Mstellar/\msun)<9$] have fractions higher with respect to the remaining, higher-mass galaxies, however we do not have a statistically significant sample in the low-mass end to confirm this trend. Also, the fractions of young stars in galaxies with $\log_{10}(\Mstellar/\msun)>9$ is low, with a mean around 0.2\% and not exceeding values of 0.3\%.

Finally, for the single case of NGC\,1272, we set $T_{\rm UV}=0$\,K and we try to reproduce the UV emission by modeling the SFH in \texttt{CIGALE} with a delayed exponential SFH in addition with a post-starburst event. We let the age of the starburst vary among 1, 10, 50, 100\,Myr, 
with a SF timescale of 1, 5, 10, 50\,Myr and a fraction in mass of 1\% or 10\%. We found that a post-starburst SFH is also not capable of reproducing the FUV emission in our target galaxies (right panel of Fig. \ref{fig:chi_square}).

\subsection{SED fitting results when the central regions are not masked}
\label{subsec:no_mask}
\begin{figure}[h!]
\centering
\includegraphics[scale=0.5]{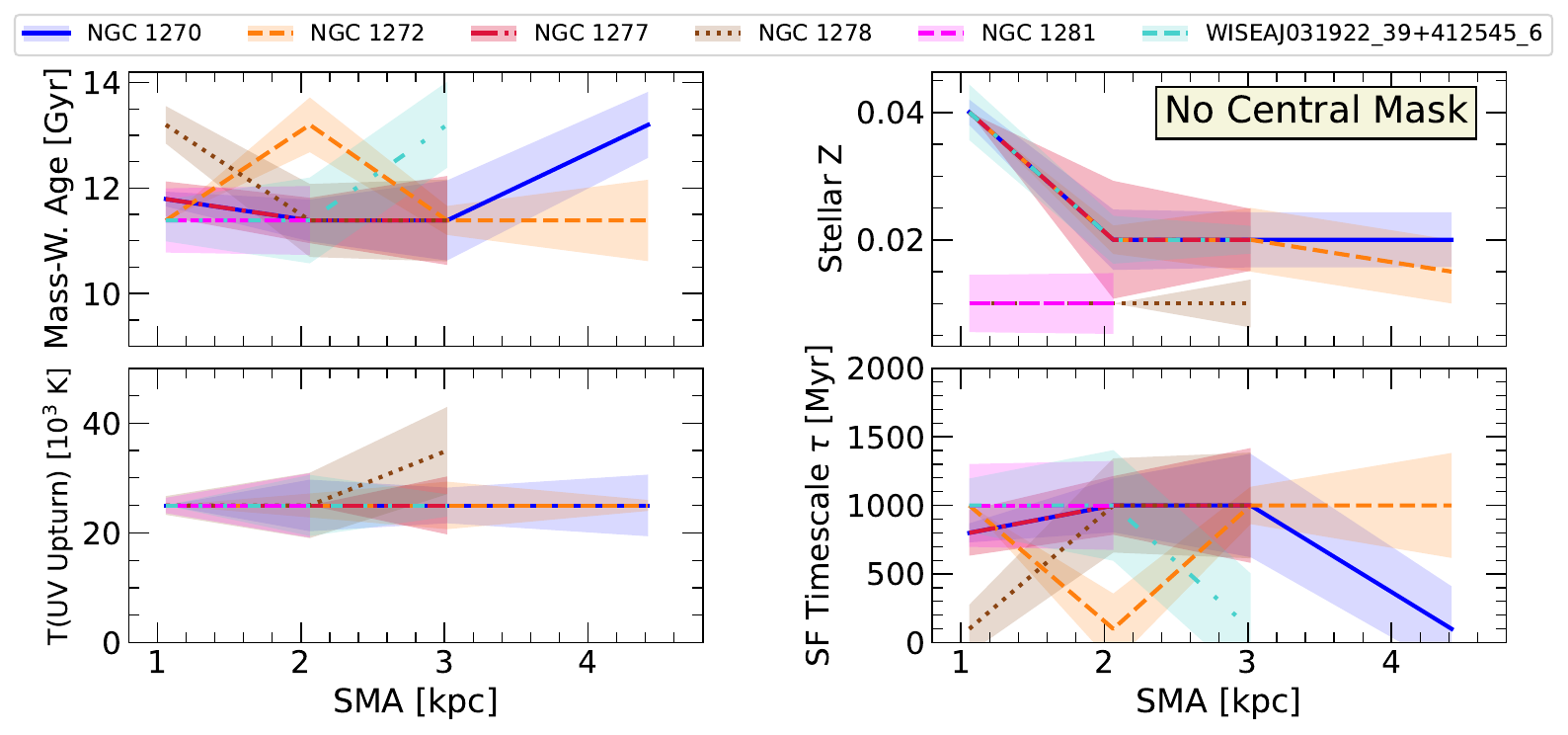}
\caption{As in Fig.~\ref{fig:res_sed}, but when the central regions are not masked.}
\label{fig:res_sed_nomask}
\label{LastPage}
\end{figure}

Figure~\ref{fig:res_sed_nomask} shows the result of the radial SED fitting for the properties of the galaxies analysed radially, when no central masking is applied. A few changes might be visible also in the outer regions, this is due to the nature of the Monte Carlo fitting technique (see \citealt{martocchia25}). For simplicity we only show here the figure of the radial profile. However, we also re-did Figs.~\ref{fig:integr_results_prop} and \ref{fig:integr_results} when no central mask is applied and the results of the paper stay unchanged. The average upturn temperature is indeed $\langle T_{\rm UV, rad, no mask}\rangle=25\,000\pm3100$\,K, while the mean SF timescale is $\langle\tau_{\rm rad, nomask}\rangle = 805\pm274$\,Myr. Concerning the integrated sample, given that the masking affects only a few objects, this does not affect the final mean results. The mean upturn temperature when no mask is applied is $\langle T_{\rm UV, int, no mask}\rangle=33\,660\pm8400$\,K, while the mean SF timescale is $\langle\tau_{\rm int, nomask}\rangle = 660\pm380$\,Myr.

\end{document}